\documentclass[12pt,preprint]{aastex}
\received{2007 July 24}
\begin{document}
\slugcomment{Accepted for Publication in The Astronomical Journal}
\title{Deep ACS Imaging in the Globular Cluster NGC 6397: The Cluster Color Magnitude Diagram and Luminosity Function$^1$}
 
\author{Harvey B. Richer\\
\affil{Department of Physics and Astronomy, University of British Columbia, Vancouver, BC, Canada\\
Email:{\tt richer@astro.ubc.ca}}}

\author{Aaron Dotter\\
\affil{Department of Physics and Astronomy, Dartmouth College, Hanover, New Hampshire\\
Email:{\tt aaron.l.dotter@dartmouth.edu}}}

\author{Jarrod Hurley\\
\affil{Center for Astrophysics and Supercomputing, Swinburne University of Technology, P. O. Box 218, VIC 3122, Australia\\
Email:{\tt jhurley@amnh.org}}}

\author{Jay Anderson\\
\affil{Department of Physics and Astronomy, Rice University, Houston, Texas\\
Email:{\tt jay@eeyore.rice.edu}}}

\author{Ivan King\\
\affil{Department of Astronomy, University of Washington, Seattle, Washington\\
Email:{\tt king@astro.washington.edu}}}

 \author{Saul Davis\\
\affil{Department of Physics and Astronomy, University of British Columbia, Vancouver, BC, Canada\\
Email:{\tt sdavis@astro.ubc.ca}}}

\author{Gregory G. Fahlman\\
\affil{HIA/NRC, Victoria, BC, Canada\\
Email:{\tt greg-fahlman@nrc-cnrc.gc.ca}}}

\author{Brad M. S. Hansen\\
\affil{Department of Physics and Astronomy, University of California at Los Angeles, Los Angeles, California\\
Email:{\tt hansen@astro.ucla.edu}}}

\author{Jason Kalirai\\
\affil{Department of Astronomy, University of California at Santa Cruz, Santa Cruz, California\\
Email:{\tt jkalirai@ucolick.org}}}

\author{Nathaniel Paust\\
\affil{Space Telescope Science Institute, Baltimore MD\\
Email:{\tt npaust@stsci.edu}}}

 \author{R. Michael Rich\\
\affil{Department of Physics and Astronomy, University of California at Los Angeles, Los Angeles, California\\
Email:{\tt rmr@astro.ucla.edu}}}


\author{Michael M. Shara\\
\affil{American Museum of Natural History, New York City\\
Email:{\tt mshara@amnh.org}}}

 \shortauthors{Richer {\it et al.}}
\righthead{CMD and LF NGC 6397}

\begin{abstract}
We present the CMD from deep HST imaging in the globular cluster NGC 6397. The ACS was used for 126 orbits to image a single field in two colors (F814W, F606W) 5$\arcmin$ SE of the cluster center. The field observed overlaps that of archival WFPC2 data from 1994 and 1997
which were used to proper motion (PM) clean the data.  Applying the PM corrections produces a remarkably clean CMD which reveals a number of features never seen before in a globular
cluster CMD.
In our field, the main sequence stars appeared to terminate close to the location in the CMD of the hydrogen-burning limit predicted by two independent sets of stellar evolution models.  The faintest observed main sequence stars are about a magnitude fainter than the least luminous metal-poor field halo stars known, suggesting that the lowest luminosity halo stars still await discovery. At the bright end the data extend beyond the main sequence turnoff to well
up the giant branch. A populous white dwarf cooling sequence is also seen in the cluster CMD. The most dramatic features of the cooling sequence are its turn to the blue at faint magnitudes as well as an apparent truncation near $F814W = 28$.  The cluster luminosity and mass functions were derived, stretching
from the turn off down to the hydrogen-burning limit. It was well modeled with either a very flat
power-law or a lognormal function. In order to interpret these fits more fully we compared them
with similar functions in the cluster core and with
a full N-body model of NGC 6397 finding satisfactory agreement between the model predictions and the data. This exercise demonstrates the important role and the effect that dynamics has played 
in altering the cluster IMF.

\end{abstract}

\keywords{globular clusters: individual (NGC 6397) -- stars:  Population II, low mass, luminosity function, mass function, white dwarfs -- Galaxy: halo}

\section{Introduction}
NGC 6397, located in the southern constellation Ara, was discovered by Abbe Nicholas Louis de la Caille during
his 2 year sojourn at the Cape of Good Hope in 1751-1752. In his original catalogue it bears the name
Lacaille III.11. It was the twelfth such cluster discovered. The cluster is located at an RA of $17^h40^m41^s$ and a Declination of 
$-53\arcdeg40\arcmin25\arcsec$ corresponding
to a Galactic longitude of $338.17\arcdeg$ and a Galactic latitude of $-11.96\arcdeg$.

The reddening in the direction to NGC 6397 seems to be well constrained with most values centered
on $E(B-V) = 0.18$. Gratton et al. (2003) derive $0.183\pm0.005$ from a careful comparison of observed color-temperature relationships for field subdwarfs (presumed to be unreddened) with those
of cluster stars. This was done in two independent color systems and the number quoted above is
an average of the two. From a wide variety of sources, Harris (1996) quotes 0.18 while an analysis
of the maps of Schlegel et al. (1996) yields 0.187 (Gratton et al. 2003). From the cluster white dwarfs (WDs) Hansen et al. (2007) derive $  E(F606W - F814W) = 0.20\pm0.03$.

The cluster distance from the Sun has recently been estimated by fitting field subdwarfs to the main sequence (MS) of NGC 6397. Reid and Gizis (1998)
used extremely low metallicity subdwarfs from HIPPARCOS to obtain a true distance modulus of
$  (m-M)_0 = 12.13\pm0.15$, while Gratton et al. (2003) used a similar technique to derive  $12.01\pm0.08$ (using $  A_V/E(B-V) = 3.1$). Using the cluster WDs, Hansen et al. (2007) obtain $12.03\pm 0.06$.  This places NGC 6397 as the second nearest globular cluster to the Sun after M4 ($ (m-M)_0 =  11.18$; Liu and Janes 1990, Dixon and Longmore 1993, Peterson et al. 1995,  Richer et al. 1997),
but note that the two clusters have similar apparent $V$-band distance moduli (12.51 for M4 and
12.63 for NGC 6397 using the Hansen et al. value). 

The cluster is known to be metal poor, but there is some disagreement on its exact metallicity and
$\alpha$-enhancement.
 Gratton et al. (2003) obtain $  [Fe/H] = -2.03\pm0.05$, $\alpha/  [Fe/H] = 0.34\pm0.02$ and
 $  [M/H] = -1.79\pm0.04$. Kraft \& Ivans (2003) suggest a value of $  [Fe/H] = -2.02 \pm0.07$ based on an analysis of FeII lines using the MARCS models \cite[]{gus75}.

Using much of the data quoted above, the age of NGC 6397 has been estimated by either fitting full isochrones,
or just the luminosity of the main sequence turnoff (MSTO). The latter technique has often been adopted as it is
notoriously difficult to correctly predict the color of the MSTO which is very sensitive to the adopted physics (D'Antona 2001).
Gratton et al. (2003) provide one of the most recent efforts in this direction and fit to the luminosity of the MSTO, deriving an age of $13.9\pm1.1$ Gyr. Allowing for some gravitational settling
in the stellar models reduces this by about 0.5 Gyr to 13.4 Gyr (Chaboyer et al. 2001). Twarog \& Twarog (2000) used data and isochrones in the Stromgren photometric system to derive a cluster age of 12.0 $\pm$0.8 Gyr, significantly younger than Gratton et al.
 In an entirely different approach, Pasquini et al. (2004)
use the abundance of beryllium in two NGC 6397 MSTO stars as a chronometer. The idea here is that
beryllium is made by spallation in the atmosphere of the star and its abundance should therefore reflect
its age. Using a model for the general enrichment of the Galaxy and a new measurement of  log$ (Be/H)$ of -12.35, they show that NGC 6397 was formed
0.2--0.3 Gyr after the onset of star formation in the Galactic halo.  If we take reionization to have commenced 13.5 Gyr ago (Spergel et al. 2003) and the Galactic halo to have formed 
$\sim$0.5 Gyr after this event, Be dating of NGC 6397 then implies an age near 12.7 Gyr.

NGC 6397  has a ``collapsed'' core (Djorgovski \& King 1986)
  and exhibits mass segregation (King et al. 1998, De Marchi et al. 2000, Andreuzzi et al. 2004). The cluster present day mass function (PDMF) was earlier measured in a single field
from HST WFPC2 data by King et al. (1998) who showed that it was quite flat between $\sim$0.2 and 0.1$M_{\odot}$ and appeared to drop steeply
below this down to the limit of the data. This conclusion should be tempered somewhat as there were
no stars below $F814W = 24.5$ in their data set. In the cluster core there is strong evidence that the stellar population has been modified by dynamical processes. At least 16 BY Dra stars have been identified in the core (Taylor et al. 2001) as well as a sample of He white dwarfs (Cool et al. 1998). In a deep Chandra exposure a total of $\sim$20 cluster sources were identified including
9 CVs, a millisecond pulsar and a collection of BY Dra stars (Grindlay et al. 2001, 2002).  

The PDMF in a globular cluster is a combination of the initial mass function (IMF), stellar evolution, and its subsequent dynamical modification. It is generally difficult to disentangle the details of the cluster evolution as neither
the complete dynamical history nor the IMF are known. In this paper we make the
most comprehensive attempt at this to date by (a) comparing with a sophisticated N-body simulation of the cluster's dynamical evolution,
(b) determining its PDMF at $\sim 2$ half light radii and in the core so that its current dynamical state can be determined (for
comparison with the modeling), and (c) attempting to determine the IMF from the cluster white dwarfs (WDs) for stars more massive than the current MSTO. 
The latter was also done in Messier 4 (Richer et al. 2004), but 
the depth of the data only allowed for a small extension of the PDMF -- up to $\sim 1.3M_{\odot}$ from the
present MSTO value of 0.8$M_{\odot}$. In NGC 6397 the coolest WDs that can be counted with confidence have evolved from MS stars with a mass of $  \sim 1.6 M_{\odot}$.
 Armed with the dynamical model to refer to, we show that,  to better than a factor of 2, we are seeing
 the number of WDs expected from the original IMF modified by dynamics and stellar evolution.
  
\section{Deep Imaging in NGC 6397}
\subsection{The Observations}

 The motivation to obtain deep HST imaging in NGC6397 comes
   primarily from its proximity to the Sun, which makes it the
   only cluster other than M4 for which the bottom of the MS 
   and WD cooling sequence are comfortably within reach of HST.
  Other compelling reasons to explore this cluster include its low metal abundance and collapsed core. 
 This globular cluster provides the best opportunity to test the white dwarf cooling age in such a metal-poor system and at the same time provide a comparison with the more metal-rich cluster, {M4}, which we recently observed with HST (Richer et al. 2002, 2004, Hansen et al. 2002, 2004). 
  Up until the present work, the Pop II hydrogen-burning limit had not been identified in any stellar population (see Richer et al. 2006).  In addition, NGC 6397 has a collapsed core, in contrast with M4, and the distribution of its various stellar members throughout the cluster have almost certainly been modified by dynamical processes. As we discuss in the following sections, we are using N-body simulations specifically developed for this cluster to understand these modifications. Among the dynamical questions we attempt to answer are: 1) do we see evidence for modification and redistribution of stellar populations in the cluster due to dynamical evolution? and 2) do these observations agree with the model predictions?

In the current program, we imaged
a single field in the cluster with HST/ACS for 126 orbits. The field was located 5$\arcmin$ SE of the
cluster core and overlaps pre-existing WFPC2 images which we use to proper motion clean the data. However, these had only a 60\% area overlap with the ACS field and were much shorter in exposure time (3.96 and 7.44 ksec in each of 1994 and 1997 compared to 179.7 ksec with ACS in 2005) so that their
utility (particularly for the faintest stars) was limited. This is not much of an issue in this paper where
we are dealing with the cluster MS as the MS LF has very few stars fainter
than $F814W = 25$.

Figure 1 illustrates the locations of our fields. The dotted lines cross at the cluster center and here faint outlines of the WFPC2
footprint indicate the locations of archival HST data. The current WFPC2 data (which were done in parallel with the ACS imaging) are indicated
by the heavy WFPC2 outline. At 5$\arcmin$ SE of the core the heavy ACS footprint shows the location of our deep outer field, with the myriad of WFPC2 outlines indicating the position of the earlier data which
is used for proper motion cleaning. Further details are contained in the figure caption.

In each orbit, two F814W ACS images and a single ACS F606W exposure were secured. We chose
this 2/1 ratio as the main scientific aim of the program was to characterize the very coolest white dwarfs
in the cluster. These are expected to have near IR deficiencies in their flux due to collisional induced
absorption in molecular hydrogen (Hansen 1999; Saumon \& Jacobson 1999; Bergeron et al. 1995a)
so we expected them to be fainter in F814W than in F606W even though they are quite cool ($\sim$4000K). 
The F606W exposure was taken during the darkest part of the orbit in order to keep the sky background
to a minimum. The typical
exposure times for these images were 750 secs with small dithers between exposures. Short exposures of 1, 5, and 40 secs were also
taken during a few orbits in order to obtain photometry of the brightest cluster stars. The roll
angle of the telescope was controlled so that the core of the cluster was imaged during the simultaneous exposures taken with WFPC2. With the WFPC2 on the core,
we obtained 3 exposures per orbit, one in each of F336W, F606W and F814W with typical exposures
of 600 secs. The F336W exposure was always taken during the darkest part of the orbit. 

\subsection{Reducing the Data}
The ACS data were reduced independently at Rice University by J. Anderson and at UBC by J. Brewer and H. Richer.
For the ACS data, the team adopted the photometry produced at Rice.  Here, we briefly outline the ACS reductions.  The details can be
    found in Anderson et al. (2008).

We arrived at a cluster sample through a two-stage
        process:  we first identified everything in the image
        that could possibly be a star, then we subjected this
        list to several tests designed to sift out the non-stars.
 After much experimentation, we determined that the faintest MS stars and WDs could best be
found by searching in F814W alone.  Our finding algorithm was based on
the fact that, due to the under-sampled WFC PSF,  faint stars will influence at most one or two pixels in a
given exposure, generating a peak (a local maximum) in some number of
the exposures.  These peaks will look just like noise fluctuations, except that they will occur in the same place in the field. So, we identified possible stars wherever a significant
number of exposures detected a peak at the same place in many images.
There were 48,785 such objects found.  Many of these are true stars, but
some galaxies and PSF artifacts also satisfied the finding criteria.

For each source, we next extracted a 5 $\times$ 5 raster from each
exposure and analyzed simultaneously the 25 $\times$ 252 = $6\,300$ F814W
and $3\,150$ F606W pixels to derive a single position, F814W and F606W fluxes, and two shape
parameters.  To purge all but the stars from the list, we removed all
the sources that:\ (1) were noticeably broader or narrower than a PSF, (2)
were so close to brighter stars as to be likely PSF artifacts, or (3)
were on linear features, and thus were likely to be bumps on diffraction
spikes.  This left us with 8,357 true stars.

\subsection{The CMDs}

Before discussion of the cluster CMD, it is important to be sure that the HST ACS data are well calibrated. We calibrated as described in Sirianni et al. (2005) but in order to carry out a sanity check on this calibration we compared our CMD with those of two others on this cluster; a well calibrated ground-based one and the
core field of the ACS Survey of Galactic Globular Clusters (ACS GGC Survey; Sarajedini et al. 2007). For the
former we used data taken from G. Piotto's web site which are in Johnson/Cousins ($V, I$). We transformed these to  ACS 
 magnitudes (F606W, F814W) using the relationships in Sirianni et al. (2005). Although the data sets have no stars in common, an over-plot showed excellent agreement with no obvious systematic differences.  Comparison with the ACS GGC Survey core
 field of NGC 6397 showed a small 0.01 magnitude difference in $(F606W - F814W)$ between the two data sets (our sequence lies redward of the ACS GGC Survey sequence). Disagreement between the two data sets at only the 0.01 magnitude level is entirely acceptable.  Since it is not obvious which data set contains this small systematic error (if it is systematic - the culprit could even be differential reddening), we remained with our own calibration.
 
 \subsection{The Uncleaned CMD}
 
Figure 2 displays the NGC 6397 CMD for 
all objects that are found as local maxima in at least 90 of 252 F814W images.  
To guard against
    including PSF artifacts, we further insisted that each source be 
    the most significant source within 7.5 WFC pixels.
 There are a total 48,785 objects in this CMD.  Even without image classification or proper motion cleaning we see a sparse giant branch which leads down to the MSTO, and then extends down the MS to faint objects with a possible termination in a cloud of field stars near $F814W = 24$. The most interesting feature of
the diagram, however,  is the WD cooling sequence which begins at about $F814W = 22.5$, shows 
an increasing concentration below about $F814W = 26$ and then exhibits a blueward 
turn below $  F814W = 27.5$. There is a possible truncation of the sequence fainter than $  F814 W \sim 27.7$ and bluer than $  (F606W - F814W) \sim 0.9$. If the WD sequence continues fainter and bluer than this, it is not possible to tell from this diagram as the strong contamination from faint blue galaxies and unidentified PSF artifacts
is enormous and swamps any subtle CMD features.  

In addition, there are a number of interesting features in this figure which are caused by the field population. The wall of field stars near $  (F606W - F814W) = 0.7$
is due to the MS turn off of the Galactic halo. Note that it is about 0.05 magnitudes {\it redder} than the cluster
turn off. This means that the field is either more metal-rich than the cluster or older. Using the Dotter et al. (2007) models, fixing both the ages and $\alpha$--enhancement  of the field and cluster to be the same, and  attributing the difference to $[Fe/H]$ alone, we derive a mean $  [Fe/H] = -1.5$ for the halo. On the other hand, if $\alpha$
and $[Fe/H]$ are held constant at the cluster value, the field would have to be about 5 Gyr older than the cluster. Since the
cluster age is near 11.5 Gyr (Hansen et al. 2007), this latter scenario seems unlikely.
Another feature of note is the drop in density  of the background below $  F814W \sim 21$. We take this 
as a sharp decrease in density of the Galactic halo and note that this occurs about 5 magnitudes fainter than the MSTO. This corresponds to a factor of 10 in distance, placing the location of the change at about 25 kpc from us (as the cluster is located at $\sim 2.5$ kpc from the Sun). With the Galactic center at about 8 kpc, this gives the galacto-centric extent
of the halo out to this distance to be about 17 kpc. This corresponds reasonably well with the extent of the {\it inner} halo of the Galaxy discussed in Carollo et al. (2007).

\subsection{The Cleaned CMD}

Figure 3 is a cleaned version of Figure 2. In this CMD we plot all the sources that pass our tests
indicating that the object is stellar. There is little doubt that 
some unresolved faint blue galaxies remain in the sample bluer than $  (F606W - F814W) = 1$ and
fainter than $  F814W \sim 27.5$ (see Hansen et al. 2007 for more details).
The WD cooling sequence is now seen clearly
to extend from $F814W \sim22.5$ down to about 26.0, below which the numbers increase rapidly,
and on to $F814W = 27.2$ where the sequence executes a hook to the blue. The sequence appears to be largely truncated at $F814W \sim 27.7$. The scatter to the
blue of the sequence (bluer than $  (F606W - F814W) = 1$) is largely due to faint unresolved blue 
galaxies.  

The diversion to the blue for the very faintest white dwarfs has been predicted by several groups (Bergeron et al. 1995a; Hansen 1999; Saumon \&
Jacobson 1999) and is likely caused by collision induced absorption (CIA) in the dense, cool
atmospheres of these stars (Zheng \& Borysow 1995, Borysow et al. 2001). This arises because hydrogen molecules can form in these high-pressure atmospheres. Being symmetric, H$_2$  does not have a dipole moment and will only weakly absorb radiation on its own via quadropole or higher transitions. A dipole moment may be induced, however, during an H$_2$--H$_2$, H$_2$--H, or H$_2$--He collision (Gustafsson \& Frommhold 2003), allowing for CIA. The stars continue to get bolometrically fainter and cooler as they evolve on the blueward track in the CMD known as the Òblue hookÓ. The strong CIA greatly suppresses the red flux which is then effectively re-radiated in the blue where the opacity is lower. The spectra of these cool white dwarfs thus departs significantly from that of a blackbody. The length of the blue hook depends partly on the age of the cluster but it is also affected by the number of massive MS stars originally present in the cluster (the high mass end of the initial mass function), and by photometric errors at these faint magnitudes. This is the first time that this theoretically predicted CIA feature has been observed in the CMD of any star cluster.  

\subsection{The Proper Motion Cleaned CMD}

 In Figure 4 we present the proper motion diagram for our data. The displacements between the 2005 ACS data and the WFPC2
          images taken in 1994 and 1997 were measured with respect to
          member stars, thus the zeropoint of motion is moving with
          the cluster.
  All motion
 has been scaled to a 10 year baseline. Unfortunately there is only a 60\% overlap between
 the earlier and recent ACS data, and, as well, the WFPC2 data are not nearly as deep or homogeneous. This limits
 the quality and depth of the proper motion selection. For example, as we will see, this does not
 allow us to investigate the faint, very blue end of the WD cooling sequence. In the original HST proposal
 we were scheduled to receive an additional 12 orbits in F814W in the spring of 2007. This would
 have allowed full coverage of our field and would have penetrated deeper than the existing WFPC2
 data. With the failure of ACS in January 2007, we were required to use the less-than-optimal archival WFPC2 images.
 
 The solid line in this diagram is a $2\sigma$ cut in the proper motion values; all stars to the left of this line are assumed  to be cluster members. The dispersion in the proper motions is due to a combination of measurement error and the cluster's
 internal proper motion.
   It is clear from this diagram that stars
fainter than $F814W = 27.5$ do not have well-measured proper motions. This is due in large part to the 
lower sensitivity and smaller
         integration time of the archival WFPC2 data.
The larger errors in proper motion measurement 
for the brightest stars is due to their near saturation. The large diffuse clump of objects centered near 
3 ACS pixels represents the field stars seen in the direction of NGC 6397. The CMD for these objects can be visualized from Figure 3 by mentally removing the cluster.

 Figure 5 is the proper motion cleaned version of Figure 3. The 2317 selected stars, their coordinates and photometry are listed in Table 1. As mentioned above, we lose about 40\% of the 
 cluster stars because of the modest overlap between the ACS and the earlier WFPC2 images and, in addition,
 we also lose in depth because of the rather short exposures with WFPC2 from 1994 and 1997. Nevertheless, the major
 features seen in Figure 2 remain and are much cleaner in this diagram. 
 
 The giant branch, MSTO, and entire MS are now cleanly delineated. Also, by comparison
 with the location of the horizontal branch (HB) in ground-based CMDs, we clearly have two blue HB stars in this diagram.  MS--MS binaries make their presence known as a scatter above the main
 sequence, the amount of scatter above the sequence being a function of the mass ratio of the binary,
 reaching a maximum of 0.75 magnitudes for equal mass systems. MS--WD binaries are found below the MS in the direction of the WD cooling sequence. There are a small number of possible binary systems
of this sort apparent in the diagram. The frequency of
 binaries in this field will be explored in detail in Davis et al. (2008, in preparation), but suffice it to say here that the frequency is low,
 in the range of 1--2\%. This is an important conclusion as Hurley et al. (2007) have recently
 shown that the binary fraction at several half light radii does not change appreciably 
 with time, so that the primordial value is more or less preserved at this large radius. By contrast, Hurley et al.
 predict that the core frequency should rise by about an order of magnitude over a Hubble time which is just what we found
 in Davis et al. (2008). If the primordial binary frequency is so low, it has important consequences for
 the formation rate of peculiar stars (blue stragglers, cataclysmic variables, novae, etc.) as well as the dynamical evolution of the cluster as a whole.
  
 The MS is
 seen to project down to $  F814W = 24$ with a sizable population, but below this the stellar
 density in the CMD drops off remarkably, so much so that there are no obvious MS
 stars fainter than $  F814W \simeq 26$, $  (F606W - F814W) \simeq 4$. This is not caused by incompleteness 
 as can be judged by Figure 3 where there is a sizable population of field stars at these colors and magnitudes (see also Richer et al. 2006). The most likely explanation for this effect is that the 
 mass-luminosity relation becomes much steeper below $F814W = 24$ so that a very small change
 in mass makes a large change in luminosity (Piotto, Cool \& King 1996; King et al. 1998). We explore this in more depth in $\S 3$. As we pointed out in Richer et al. (2006), this change in the mass-luminosity relation arises
 because at the low masses of these stars, electron degeneracy pressure is important
 in supporting the star against collapse. This means that the central temperature of the
 star no longer increases with pressure as it does for one supported by a classical gas. 
 This dependence on degeneracy pressure will decrease the central temperature at lower masses and thus the nuclear energy generation rate which
 still supplies most of the luminosity of the star (Kumar 1963, Chabrier and Baraffe 2000, Kumar 2002). 
 
 In addition to the apparent faint end of the cluster MS, there are three stars that scatter
 away from an extension of the sequence. These are possible 
 interlopers from the extensive field population,  or stars with poorer photometry. Inspection of these stars on the images shows that they are all on or near diffraction spikes or bright stars. There thus does not appear to
 be any MS population fainter than about $F814W \simeq 26$ and redder than  $  (F606W - F814W) \simeq 4.0$.
 We have suggested in Richer et al. (2006) that this could be the termination point of the hydrogen-burning
 MS for these metal-poor stars. In Figure 6 we have included two theoretical
 locations for the termination of hydrogen-burning in low-mass metal-poor stars. These are taken
 from Baraffe et al. (1997, filled square) and Dotter (this paper, filled circle). The general
 agreement between the theoretical and (possibly) observed location of the termination of hydrogen
 burning in these stars is good. Both of these sets of models suggest that the mass at
 this termination point is $0.083 M_{\odot}$. We caution, however,  that other models suggest quite
 different
 results. For example, Montalban et al. (2000) would place the hydrogen-burning  limit at  $  F814W = 29.32$ in NGC 6397. This low value is largely due to the fact that in these models the
 lowest mass star capable of hydrogen burning is $0.075 M_{\odot}$. It is interesting to note, however,
 that their model for $0.083M_{\odot}$ predicts an F814W magnitude of 26.87, in reasonable agreement
 with those of the other two groups mentioned above.  In addition, there is also no hint of any cluster brown dwarfs which could be as bright as a few magnitudes dimmer than the least massive star capable of burning hydrogen (Burrows et al. 1997).  
   
  The stars in the ``blue hook'' region of the WDs are seen to be moving with
the cluster, so for the first time we are definitely seeing this feature in a cluster CMD. The depth
of the diagram has allowed us to observe, as well,  an apparent truncation in the sequence at about $F814W = 27.7$.
A detailed discussion of the cluster WDs is given in Hansen et al. (2007), but we note here that this truncation has allowed for a uniquely precise age for this cluster to be determined from WD cooling.
In an upcoming paper (H. Richer et al. 2008, in preparation) we develop the concordance age for NGC 6397, that is the age determined from the joint probability 
distribution of  the WD cooling age and the MSTO age.  
  Both these techniques have their own difficulties.
Modeling globular cluster MSTOs and reddening  and distance related issues will continue to be a concern even if distances are derived to 
a handful of clusters by the Space Interferometry Mission. In addition, uncertainties in some of the input physics (such as opacities, gravitational settling, nuclear reaction
rates, and particularly the mixing length parameter) will continue to be a source of systematic error
in MSTO age determinations. Typical $1 \sigma$ errorbars from MS fitting are currently ${\pm1}$ Gyr (eg Gratton et al. 2003, Salaris 
\& Weiss 2002).
By contrast, WD  
cooling ages are derived from the overall distribution of WDs in the CMD (Hansen et 
al. 2004, 2007). They are less sensitive to distance and reddening uncertainties, but have their own set of difficult physics. Among these are neutrino losses, sedimentation of carbon and oxygen, crystallization and modeling collision induced opacities. The important point here is that the physics entering into models of cooling
WDs is generally different from that input into MS models so that ages in a given
cluster determined with both techniques can provide some estimate of these systematics.

\subsection{A Digression: Undiscovered Faint Stars in the Galactic Halo}

At this point we can legitimately ask whether we are seeing fainter stars in NGC 6397 than are known in the Galactic halo. We have made the claim above that we have run out of stars along the cluster
MS and have penetrated to the hydrogen-burning limit.  This would suggest that we ought not to
see dimmer
field stars of similar metallicity in the halo.  We investigate this by over-plotting on our CMD some of the faintest known low metallicity halo
subdwarfs. To be included in this plot a star had to have 1) a known metal abundance and be less than $  [Fe/H] = -1.5$  (Gizis 1997, Phang-Boa and Bessell 2006, Woolf and Wallerstein 2006), 2) V and I photometry (so that it could be transformed to F606W and F814W), and 3) a known distance (so it could be placed at the cluster distance). This means that some recently discovered extreme subdwarfs (e.g. Gizis and Harvin 2006, Burgasser et al. 2007) were not included as the data for these are not complete enough. While we may have missed some stars in our literature 
search, the small 5 star sample should be representative. These are illustrated with open circles in Figure 6 (and their properties tabulated in Table 2) where the photometry and distances for the stars have been taken
from Reid and Gizis (2000). The colors and magnitudes of these stars were then transformed to the ACS system using the transformations in Sirianni et al. (2005).
Although the photometry and distance estimates for these stars and their transformations are somewhat uncertain, the trend is clear enough. We are seeing stars in NGC 6397 that are more than 1.5 magnitudes fainter than the lowest luminosty star in this sample (LHS 322).  
This suggests that there are still very low luminosity, cool subdwarfs awaiting
discovery in the Galactic halo. Using the models from Baraffe et al. (1997), the stars  in Table 2 are not lower in mass than $  0.091 M_{\odot}$, whereas the least massive cluster stars are $  0.083 M_{\odot}$.

We note here that the extreme subdwarfs currently being found
by various searches (eg Lepine et al. 2004) are the expected low mass end
of the halo MS. What is of great interest, of course, is to see if any low metallicity field subdwarfs (or substellar objects)
turn up that would place them fainter than the apparent termination of our NGC 6397 MS.
The suggested sub-stellar object 2MASS J05325346+8246465 discussed in Burgasser et al. (2003) is not
included in Figure 6 as it has no known distance, V photometry or metal abundance.

We can estimate roughly the relative number of unobserved faint halo stars in the following way.  
The number of MS plus giant stars from $  F814W \sim 12$ down to $F814W = 24.5$ (the magnitude the faintest 
field subdwarf in this sample, LHS 377, would have if it were a cluster member) in our NGC 6397 CMD is 2031. The number of
cluster stars fainter than this is 6. Hence 1 in about 340 cluster stars is fainter than the least
luminous field halo star shown here. We thus expect about 0.3\% of all halo stars to be fainter than LHS 377. If subdwarfs constitute 1 in $\sim 600$ of all stars locally (Siegel et al. 2002), then about
1 in $200\,000$ stars in a volume limited sample in the Solar neighborhood should be fainter than LHS 377,  one of the faintest metal-poor subdwarfs currently
known. This estimate is likely a strong lower limit as we have not accounted for the enhanced depletion of
low luminosity cluster stars via evaporation. In the dynamical model that we discuss in $\S 3.3$ about
95\% of the very lowest mass cluster stars have evaporated by 12 Gyr.

\section{The Cluster Luminosity and Mass Functions}

 The cluster luminosity function (LF) is obtained by counting the number of main-sequence stars
 as a function of magnitude. For this we use the proper-motion selected sample which makes field star contamination no longer an issue in determining the LF.  However, the LF contains limited information and cannot be compared directly from one globular cluster to another because of metallicity effects. That is, the mass of a MS star at a given luminosity depends on the metal abundance of the star. The quantity that can be compared between clusters, however, is the cluster MF which we derive below.

We reviewed the recent literature in order to obtain the best current physical parameters for
NGC 6397 to use in this analysis; these are compiled in Table 3. From the table we take as our basic parameters
for NGC 6397 a true distance modulus of $12.03\pm0.06 (1\sigma)$ and a reddening $  E(B-V) = E(F606W - F814W) = 0.18\pm0.03$.
This is the WD distance modulus (Hansen et al.  2007) and an average reddening from several
techniques.

\subsection{Previous Work}  
The MF for NGC 6397 has been determined by a number of groups recently, largely because of its  proximity
to the Sun, its extreme metallicity and its dynamical state.  The most useful of these LFs come from HST data. Early on, Paresce, De Marchi, \& Romaniello (1995) suggested that the cluster had a deficiency of low-mass stars. However, there were no earlier data from which proper motions could be used to clean the images making field star contamination at low luminosity a difficulty in this study.  King et al. (1998) used proper motions to eliminate most of the foreground/background stars and found a shallow MF that flattened out at about $0.3 M_{\odot}$ and showed a possible decline below about $0.1 M_{\odot}$.  De Marchi et al. (2000) used both NICMOS and WFPC2 data on HST to argue that the cluster MF could be best fit with a lognormal MF and that there were virtually no cluster MS stars below $   M_{F814W} \sim 12$ (corresponding to $  F814W \sim 24.4$). Richer et al. (2006) showed that the cluster
MS projected down to about the theoretical location of the hydrogen-burning limit for low metal-abundance stars near $  F814W = 26$. 

\subsection{NGC 6397 Luminosity and Mass Functions}
  Our approach in this work (as it was in M4 (Richer et al. 2004)) is to leave the observational data
  untouched and make any needed corrections in the theoretical plane. This has the advantage that the errors in the observational quantities remain as purely statistical without introducing the problem of error propagation. For example, there are corrections for completeness effects in the star counting that must be applied, and we have incorporated these within the theory.

The completeness of our data is an important issue in deriving the cluster MF. It was determined with artificial star tests, which are described in Anderson et al. (2008). Briefly, the added stars were inserted into the {\it individual} images at the same position (corrected for dither) with the appropriate flux and noise. The artificial stars were then subjected to the same finding and measuring procedures developed for the real stars. A star was
considered to be found if it lay within 0.75 pixels in $x$ and $y$ of its input position and within 0.75
magnitudes in F814W. We did not use the F606W images to compute the completeness.
Including F606W in the finding 
made it harder to develop a robust incompleteness scheme, since experimentation
found that the faintest stars were more reliably found by using only F814W.
The additional benefit of this was that the same completeness can be
used for the WD stars as the MS ones where the stars are very quickly dropping out in F606W. 
 Insisting that stars have a consistent flux in F606W would not
remove very many objects, except where the F606W flux completely
drops out.  Importantly, the completeness goes to zero at a magnitude well
above where we can still measure stars (by finding them only in F814W).
 We find stars several magnitudes fainter than the F606W drop-out level assuming
that we can either determine membership from proper motions, or just
assume membership because there shouldn't be anything else so red. Because of the many orientations, partial coverage, 
and lack of depth  of the WFPC2
data used for proper motion cleaning, it proved to be virtually impossible to carry out successful experiments which incorporated
incompleteness corrections to these earlier epoch images. However, we examined what the effect might be on the lower main sequence
by comparing the number of stars with reliable proper motions to that of all stars as a function of magnitude.
Most of the faint stars in this test were field stars.
This ratio was constant from $  F814W = 21$ down to 24 at about 64\%, from $24 - 25$ it dropped by  10\%
and from 25 through 26 it dropped by an additional few more percent. 
The drop in the observed density of MS  stars is so drastically larger than this that it cannot be due to
 incompleteness or to our inability to measure proper motions, but is a real deficiency of stars at these faint magnitudes in the cluster. We have therefore not made any additional corrections for proper motion incompleteness. 
Table 4 contains the (inverse) corrections from the ACS images that we applied to the theoretical LFs.

 Another point to be made about deriving the MF is that the theoretical models used here, those of Dotter et al. (2007 with additional models between 0.083 and 0.1 $M_{\odot}$), which contain
 the most up-to-date physics in any generation of models, do not fit the lower MS (fainter than $  F814W \sim22.5$) of NGC 6397 particularly well, although the fit is better than any previous set of models. We illustrate this 
in Figure 6 where a metal-poor model ($  [Fe/H] = -1.9$, $  [\alpha/Fe] = + 0.4$) is overlayed on the cluster CMD using our adopted
distance modulus and reddening ($  (m-M)_{F814W} = 12.36$; $  E(F606W - F814W) = 0.18$). The agreement is quite
reasonable down to $F814W = 22.5$ and then for about the next 1.5 magnitudes the isochrone is either too blue or too low in luminosity. This is likely due, in part, to low mass models being less luminous at a given mass than real stars. This phenomenon has been known for 
some time (Chabrier \& Baraffe 2000).  The discrepancy may also be due deficiencies in the color-temperature relationship, most probably in the F606W band where molecular
absorption is more severe and where any metallicitiy mismatch between the real stars and the models is large (Chabrier \& Baraffe 2000). This may be the same problem in producing theoretical spectra of metal-poor cool stars in the V-band and the greater success in the redder bands (Bedin et al. 2001; Delfosse et al. 2000). To somewhat mitigate any effects of difficulty in modeling the F606W LF, we derive the cluster LF using F814W.

From the proper-motion selected cluster MS we derived the F814W LF by binning the stars in 0.5 magnitude intervals.
 We then used the isochrones of Dotter et al. (2007) with extension to the H-burning limit, along with power-law and lognormal MFs, to generate theoretical LFs. These LFs were multiplied by the detection efficiency in the data so that they could be directly compared with the observations.
The theoretically derived LFs were normalized over the entire magnitude range and this normalization is thus a parameter in the fit.  
 The best-fitting model LF was chosen via a $\chi^2$ test between the observed and theoretical LFs.  The $\chi^2$ analysis included variations in the distance, age, and IMF parameters: either slope for the power-law IMF or central value and dispersion for the lognormal IMF.
The MF of the best fitting LF was taken as the appropriate cluster MF.

Figure 7 illustrates the observed LF in F814W along with the two best-fitting theoretical LFs: one with a power-law MF and the other with a lognormal function. The power-law is described by the slope $\alpha$ defined as $\Phi(M) \propto M^{-\alpha}$ where $M$ is the mass and the Salpeter (1955) value of $\alpha$ is 2.35. The lognormal MF (Miller and Scalo 1979, Chabrier 2003) is characterized by a Gaussian distribution $\Phi(M) \propto e^{-[log (M/M_c)^2]/2\sigma^2]}$ where $M_c$ is the mean mass of the distribution and $\sigma$ is the mass dispersion.  The data are shown with their associated $\sqrt n$ error bars in 0.5 magnitude non-overlapping bins.
The small bump at bright magnitudes in the LF ($F814W = 18$) is caused by the onset of molecular absorption in the stellar atmosphere near a mass of $0.5 M_{\odot}$ (Chabrier and Baraffe 2000, Chabrier 2003). 
This changes the slope of the mass-luminosity relation in such a way that a modest range in stellar mass creates a small range in luminosity
so that the stars pile up at this luminosity. The
large bump seen both in the data and the models near $F814W = 21$ (mass $\sim 0.35 M_{\odot}$) comes from the onset of complete convection in the star while the small bump slightly fainter than  $F814W = 23$ ($M \sim 0.12 M_{\odot}$) is caused by the onset of degeneracy in the core of the star. 

Formal $\chi^2$ fits between the data and theoretical LFs with power-law MFs, where $\alpha$ was allowed to range from -1 to +1 in 200 steps, selected $\alpha = 0.13$  ($\chi^2 = 94$ for 28 bins and the two constraints of slope and normalization) as the best fitting power-law MF (see Figure 7). This is in general agreement with earlier studies that found a rather flat MF for this general location in NGC 6397. What is of interest here is that a single power-law MF slope seems appropriate from the MSTO to just below $0.1 M_{\odot}$ ($F814W = 24$). Beyond this, to the termination of the MS  at $0.083 M_{\odot}$, the model (including incompleteness) predicts 37 stars whereas 12 were found. Whether this suggests a flattening of the slope of the MF at these very low masses, small number statistics, dynamical evolution of the cluster, deficiencies in the theoretical mass-luminosity relation at very low masses or some remaining uncertainty with the incompleteness corrections is not clear. In any case, the power-law LF appears to do a decent job of predicting the number of stars in the cluster from the MSTO to below $0.1 M_{\odot}$.

Similar tests were performed with the lognormal MF where $M_c$ was allowed to range from 0.083 to $0.8 M_{\odot}$ and $\sigma$ from 0.1 to $2 M_{\odot}$. As $\sigma$ grows increasingly large the lognormal MF begins to resemble a flat MF as the large dispersion washes out sensitivity to $M_c$. 
The lognormal MF generally produced better $\chi^2$ values than the best fitting power-law MF for $M_c = 0.083$--0.4 and $\sigma =0.75$--2. The lowest $\chi^2$ value occurred for $M_c=0.27$ and $\sigma=1.05$ ($\chi^2 = 48$ for 28 bins and 3 constraints, Figure 7). This best fit lognormal LF predicts 21 stars in the three faintest bins.  This is a good deal better than the best fit power-law (37 stars) but still nearly twice the number of real stars (12).

There is some disagreement in the literature regarding the applicability of the power-law and lognormal MFs. We have shown here that, in fact, a single power-law MF performs quite well down to the reliable limit of the data. The lognormal MF performs even better but it has two adjustable parameters whereas the power-law has only one. The chief improvement of the lognormal over the power-law is that it truncates the LF at the extreme low mass end. This end is subject to the most uncertainty in both the data and the theoretical models and the lognormal MF may simply make accomodations for these uncertainties.

Finally, both best fit LFs in Figure 7 incorrectly predict where the peak of the LF should occur. Since this peak is associated with the transition to fully convective stars, one possible explanation is that the models transition to being fully convective at a slightly lower mass than the real stars. Although this is still a possibility, there is evidence that the mismatch is not due to problems in the stellar evolution models but rather to the neglect of dynamical effects on the MF as we discuss below in $\S 3.3$.

ACS photometry in both F606W and F814W bands on the core of NGC 6397 from the ACS GGC Survey provides us with an informative comparison. The photometry used to make the F814W LF was not proper-motion cleaned. It was, however, improved by removing all stars that did not appear in both F606W and F814W images, removing all stars with photometric errors greater than 0.4 mag in either filter, and omitting stars that lie within a 300-pixel radius of the cluster core due to crowding concerns.  Artificial star tests provide completeness estimates that are applied to the theoretical LFs as described earlier in this section.

Formal $\chi^2$ fits were performed on this LF in the same manner as described above but with poorer results. The best fit power-law MF is $\alpha = -0.68$ ($\chi^2$=452 for 24 bins and two constraints). The best fit lognormal MF is $M_c$=0.86 and $\sigma$=1.05 ($\chi^2$=316 for 24 bins and three constraints). It was necessary to increase the possible range of $M_c$ in order to find the best fit value. The fact that the lognormal MF peaks at a significantly higher mass than in the outer field is evidence for   dynamical evolution. As in the previous case, the lognormal MF produced a better fit than the power-law MF. As expected, the core LF suggests a much more top-heavy mass distribution than the outer field due to mass segregation and binaries.

\subsection{Dynamical Influence on the Luminosity Function}

Globular clusters do not evolve quietly: their stars interact with each 
other and the gravitational field of the galaxy for billions of years.
In a companion paper, Hurley et al. (2008, in preparation) describe a
series of $N$-body simulations aimed at understanding the dynamical
evolution of NGC$\,6397$.
These simulations are performed on GRAPE-6 boards (Makino 2002) using
the {\tt NBODY4} code (Aarseth 1999).
The effects of stellar evolution, binary formation and disruption,
and the tidal effects of the Milky Way's gravitational field are
included (see Hurley et al. 2001 for details).

Unfortunately, the computational constraints of the $N$-body method mean
that direct models of globular clusters such as NGC$\,6397$ are not
currently possible -- these will have to wait for the next incarnation of 
the GRAPE special-purpose hardware.
In the meantime much can be learnt from non-direct models that represent
the cluster in question as closely as possible.
Here we will make use of an $N$-body model that started with $100\,000$
stars and was evolved to an age of $20\,$Gyr 
(Hurley, Aarseth \& Shara 2007,  Hurley et al. 2008).
This model cluster reached the end of its initial core-collapse phase at 
an age of about $16\,$Gyr with $\sim 20\,000$ stars remaining and a 
half-mass radius of $5\,$pc (the corresponding 2-dimensional projected 
half-light radius is $\sim 3\,$pc).
Thus, the simulation provides information on the state of a 
post-core-collapse cluster but at a stellar evolution age greater than 
that of the real cluster.
It would be possible to bring forward the core-collapse time of the model
cluster by evolving it within a stronger tidal field.
This could be achieved by placing the cluster on an orbit that is closer
to the Galactic center -- the simulation utilised here had
the cluster on an orbit at $8.5\,$kpc from the Galactic center while
NGC$\,6397$ gets as close as $4\,$kpc.
However, the simulation would then have to start with a larger $N$ to
ensure that a bound cluster remained at the stellar evolution time
of interest ($\sim 12\,$Gyr in our case).
Full details of the effects of the tidal field and number of stars on the
evolution timescales of star clusters will be discussed in
Hurley et al. (2008).

For our purposes, a simulated MF can be extracted at any time and location within this simulation and compared to the data by multiplying it by our preferred theoretical mass-luminosity relation and the completeness corrections. The NBODY4 code does contain its own stellar evolution algorithm which 
  provides stellar luminosities but we effectively override this. This
  allows us to examine the effects that dynamics have on the mass function over and above quiescent stellar evolution.  We have adopted the 12 Gyr isochrone shown in Figure 6 as the mass-luminosity relation and place no constraints on the dynamical age.  The reason for doing so is that, although the dynamical simulation was prepared to approximate the behavior of NGC 6397, it is not possible to account for every significant gravitational interaction that the cluster has had over its lifetime.  
  
The dynamical simulation adopts a broken power-law IMF (Kroupa, Tout, \& Gilmore 1993) and a primordial binary fraction of 5\%.  This fraction stays roughly constant in the outer field (1 -- 3 half light radii) but rises to 10 -- 15\%  at a dynamical age of 12 Gyr and greater in the core (within 1 half-light radius).  Most, if not all, binaries in the data should be unresolved and therefore need to be treated as unresolved in the model LFs.  S. Davis et al. (2008, in preparation) find the observed binary fraction to be 1-2\% in the outer field and $\sim$15\% in the core.  As a result, the effects of unresolved binaries have been ignored in the outer field but included in the core.  Generally speaking, treating binaries as unresolved in the model LFs reduces the relative number of stars in bins fainter than the largest peak compared to brighter bins and reduces the overall number of stars by the number of binary systems.

Figure 9 shows the observed LF along with LFs derived from the adopted mass-luminosity relation and MFs taken from the dynamical simulation between 1 and 3 half-light radii.  These LFs span a wide range of {\it dynamical} ages: $\tau_{Dyn}$=0, the initial conditions; $\tau_{Dyn}$=12 Gyr, the stellar evolution age; $\tau_{Dyn}$=15 Gyr, shortly before core collapse; $\tau_{Dyn}$=18 Gyr, well after core collapse; and $\tau_{Dyn}$=20 Gyr.  The model LFs were normalized to the number of real stars in the range plotted in the figure. The data are best fit by the $\tau_{Dyn}$=18 Gyr LF.  It reproduces the correct shape of the LF over the range of luminosity plotted and also matches the location of the largest peak in the observed LF.

Figure 10 shows the same type of comparison with the observed LF in the core.
The model LFs were constructed from the same theoretical mass-luminosity relation but now with MFs of stars that lie within 1 half light radius of the center.  In all cases, the model LFs have been normalized to the number of real stars shown, those with $  F814W \geq$17. Again, the $\tau_{Dyn}$=18 Gyr LF provides the best fit to the data.
 Since we find a mismatch between the stellar evolution age (12 Gyr) and the dynamical age (18 Gyr), Figures 9 and 10 cover only those stars whose MS lifetimes are $>$ 20 Gyr. We do this to exclude the effects of stellar evolution in the dynamical simulation which are in step with the dynamical time scale.

Figures 9 and 10 indicate that the MF of NGC 6397 has been significantly dynamically modified.  NGC 6397 is observed to be in core collapse (Djorgovski \& King, 1986) and the LF comparisons indicate that the data are best fit by a post core collapse model with $\tau_{Dyn}$=18 Gyr.  Disconnecting the stellar evolution and dynamical ages and using our best mass-luminosity relationship were crucial steps in reaching this important conclusion.

Our analysis demonstrates that dynamical evolution is a necessary consideration in interpreting the present day MF of a globular cluster. Clearly, the IMF is not preserved in post-core collapse clusters either at the center or further out, as shown in Figures 9 and 10.  Hurley et al. (2008, in preparation) discuss the evolution of the cluster MF as a function of time and radius in greater detail.

If the cluster IMF were described by either a power-law or a lognormal function (as discussed in $\S 3.2$), and if we restrict our analysis to only one field, then the arguments in favor of dynamical evolution are unnecessary.  However, we can now present three strong arguments against these simple assumptions: (1) the different parameters required to fit power-law or lognormal MFs in the core and outer fields can be explained by one IMF plus dynamical evolution, (2) the location of the largest peak in the dynamical model LFs is seen to approach the observed location as the cluster evolves dynamically, and (3) the observed LF is best fit by a post-core collapse model LF which is consistent with the known dynamical state of NGC 6397.

\subsection{Extending the Main Sequence Mass Function Beyond the Turnoff}

The cluster WDs have evolved from stars originally more massive than the present MSTO. They thus sample the cluster MF to higher masses than are available using only the MS. In principle, we can exploit the WDs to extend the presently observed MS MF up to higher masses. As we showed in Richer et al. (2004), if we ignore dynamical effects and assume that the number of stars are preserved as they evolve through the various post MS phases 

\begin{equation}
\frac{N_2}{N_1}=\left(\frac{1-\alpha_1}{\alpha_2-1}\right)\left[1-\left(\frac{m_2}{m_3}\right)^{\alpha_2-1}\right]\left[1-\left(\frac{m_1}{m_2}\right)^{1-\alpha_1}\right]^{-1}
 \end{equation}
 
\noindent where we have assumed a broken power-law for the MF. Here $m_1$, $m_2$, and $m_3$ are the lower mass limit, break point, and upper mass limit of the MF, respectively; $N_2$ and $N_1$ are the number of WDs and MS stars in our field; and $\alpha_1$ and $\alpha_2$ are the MF slopes for the MS below and above the break point in the MF.
 
 We will take $m_1 = 0.5M_{\odot}$ in order to match as well as possible the dynamical effects on
 the WDs and the MS stars. For example, if we had chosen $m_1$ to be the hydrogen-burning limit at  $0.083 M_{\odot}$ then differential evaporation between the WDs and MS stars would be a concern
 in our considerations. From the data, the numbers used in this analysis (all corrected for incompleteness as they must be because the WDs suffer much greater incompleteness than the MS stars) are $N_2 = 390$,
 $N_1 = 484$, $m_1 = 0.5$ and $m_2 = 0.7 M_{\odot}$ (for this choice see below).
  We have set the limit of data for counting WDs to be $  F814W = 28.0$ (which is the 50\% 
 completeness limit), this corresponds to $  M_{F814W} = 15.6$. Taking  a WD mass of $0.6 M_{\odot}$ 
for these stars (see Figure 19 in Hansen et al. 2007) and letting the stars have pure hydrogen atmospheres, the cooling time to
 this magnitude is 10.4 Gyrs (Richer et al. 2000). We take the age of the cluster as 11.5 Gyr (Hansen 2007) so that the MS lifetime of the progenitor to this WD is 1.1 Gyr. From the
 Dotter et al. (2007) metal-poor models the initial mass of this star was $m_3 = 1.6 M_{\odot}$. This
 is the mass limit up to which we can extend the current cluster MF.
 
 Using equation 1 above we  then find that $\alpha_2$ is very large -- in the range of 5.1, implying a very steep MF.
 We note that a similar but less dramatic result was found for M4 (Richer et al. 2004).  For M4, above the break point, the MF slope needed to steepen from about 0.1 to near the Salpeter (1955) value at $\alpha_2 = 2.3$. In both cases, the derived values of $\alpha_2$ assume that $m_2$ was located at the current MSTO mass. If such a break were to occur at a higher mass, as suggested by the Kroupa et al. (1993) MF, then the derived value of $\alpha_2$ would be even larger.

It seems highly unlikely that such a large change in the MF slope should occur just above the level of the current MSTO in either cluster, let alone both. In order to interpret the result obtained with equation 1 above, we have extracted $N_2/N_1$ from the N-body model using the same range of MS masses for $N_1$ as above for the complete range of dynamical ages. Equation 1 requires completeness corrected numbers but we prefer to leave the observations untouched below and apply the completeness correction to the model numbers. These are then plotted in Figure 11. The uncorrected ratio from the observations is $N_2/N_1$ = 0.62.
 We have used MS stars only in the range of masses from $0.5 - 0.7 M_{\odot}$ and not up to $0.8 M_{\odot}$ as the cluster turnoff falls below this mass limit
for ages greater than 12 Gyr. For the MS stars we have applied a constant completeness correction of 0.93 as they are all bright stars (see Table 4). 
 
  One would expect, naively, that the ratio would continue to rise
as WDs are continuously produced. However, $N_2$ is the number of WDs that are observed and by  10 Gyr many of the cluster WDs have cooled below our observational limit (Richer et al. 2000). 
The ratio $N_2 / N_1$ in the model
rises until about 8 Gyr then falls slowly to a value near 0.35 by 11.5 Gyr (the cluster age), subsequent to which it falls more slowly so that by 18 Gyr (the age of the model that best matches the dynamical state of the cluster) the value is near 0.24. Hence at the stellar evolution age of the cluster the model ratio is about 56\% that observed and by the dynamical age it is just under 40\%.
Given the mismatch between dynamical and stellar evolutionary ages and deficiencies in the model and perhaps in WD cooling rates, it is difficult to interpret this
diagram in the light of the real cluster statistics. It may be that too few cluster stars have evaporated from the model given its input orbit. Some evidence for this can actually be seen in Figure 9 where the model appears to have too many low mass stars compared to the real cluster. 

However, what is eminently clear is that low ratios
of $N_2 / N_1$ can be obtained in a model cluster with just stellar evolution and dynamics operating
on a standard IMF. Thus, to a first approximation, there is no need to invoke drastic MF changes in the
real cluster. We have seen that an input  Kroupa IMF plus stellar and dynamical evolution can likely
explain the data adequately.

 \subsection{A Peek Into the Future of Luminosity Functions}

In the not too distant future, JWST will be operating and will be able to image objects several magnitudes
fainter than is possible with HST. The current specifications indicate that JWST will have broadband filters
at 700 and 900 nm which will be extremely useful for sampling the entire stellar population in globular clusters. With proper motion cleaning, LFs in only a single color will be a very efficient manner of extracting
information about these systems. This will substantially reduce the amount of observing time
that otherwise would be required to construct a cluster CMD. 

In Figure 12 we illustrate the complete F814W LF for our field in NGC 6397 as a function of age extracted from the simulated N-body model. The contribution from cluster MS stars is largely restricted to brighter than $F814W = 24$. Fainter than this the LF is entirely dominated by WDs. There
are interesting details at the bright end which are age sensitive, but they are largely lost here in small number statistics.
However, at the faint end, the peak in the WD LF marches dramatically fainter as the cluster ages.
As can be seen, the peak in the WD LF changes by about 1 magnitude / Gyr at ages in the range of 12--14 Gyr and will be an exquisite age diagnostic in the future. By contrast, between 10 and 14 Gyr, the luminosity of the turnoff changes only by 0.3 magnitudes, less than 0.1 magnitudes / Gyr.
  
\section{Summary and Conclusions}

In this paper we have presented results from a very deep imaging program in a single field of the globular cluster NGC 6397 with the ACS camera on HST. A brief summary of the data and reduction routines
were provided. The reductions were so novel that we are devoting a complete paper to them (Anderson et al. 2008). The CMD developed for this cluster shows, for the first time, a truncation in the white dwarf cooling sequence (at $F814W \sim 28$). This greatly facilitates determination of an accurate white dwarf cooling age (Hansen et al. 2007). The cluster MS is traced to approximately where current models
predict that the hydrogen MS should end ($F814W \sim 26$, $ \sim0.083M_{\odot}$). Beyond this limit,  there are no obvious MS cluster stars present in the CMD. If
there were fainter MS stars in the cluster we would have found them. 
 Observations in the infrared will prove to be useful here in confirming this result. The cluster MS luminosity and mass functions were explored in some detail. Fitting to a power-law MF, the slope  in our field was found to be very flat, in general agreement with earlier results. The MF was found to be considerably more top-heavy in the cluster core, in line with general perceptions about mass segregation. A novel feature of the present work was the comparison of our cluster LFs with those
 extracted from an N-body calculation meant to simulate the cluster. While this calculation was still somewhat imperfect (too few stars originally, orbit not quite matching that of NGC 6397, primordial binary frequency likely too high), the agreement 
 between the observations and model predictions was striking. In particular, model LFs at the same dynamical
 age (just after core collapse) were found to simultaneously fit data from both the cluster core and a 
 field $5 \arcmin$ from the center. Finally, we found that the faintest cluster WDs that could be
 counted with confidence (that did not suffer unduly from incompleteness) originated
 from MS stars with masses up to  $  \sim 1.6 M_{\odot}$ whereas the cluster MS turnoff is near
 $  \sim 0.8 M_{\odot}$.
    The implied number of original MS stars between $  0.8$ and $  1.6 M_{\odot}$ are, within a factor of 2, 
 consistent with the effects of dynamics and stellar evolution operating on the original cluster IMF.

  \acknowledgements

This work was supported by NASA/HST grant GO-10424 (J.A., B.M.S.H., I.R.K., J.S.K., R.M.R.), NASA through a Hubble Fellowship (J.S.K.), the U.S.-Canada Fulbright Fellowship Committee (H.B.R.), the Natural Sciences and Engineering Research Council of Canada (H.B.R.), and the University of British Columbia. We would also like to thank the Pacific Institute for Theoretical Physics (PITP) for support of this work, which was done
while A.D. was a visitor there. This research is based on NASA/ESA Hubble Space Telescope observations obtained at the Space Telescope Science Institute, which is operated by the Association of Universities for Research in Astronomy Inc. under NASA contract NAS5-26555. These observations are associated with proposal GO-10424. This paper also uses data from the ACS Survey of Galactic Globular Clusters (GO-10775; PI: Ata Sarajedini).  We thank Sarajedini and the other members of the ACS Survey team for the use of these data. MS and JH thank John Ouellette for excellent system administration of the GRAPE-6 computers at the
American Museum of Natural History.

\clearpage

\begin{figure}
\
\plotone{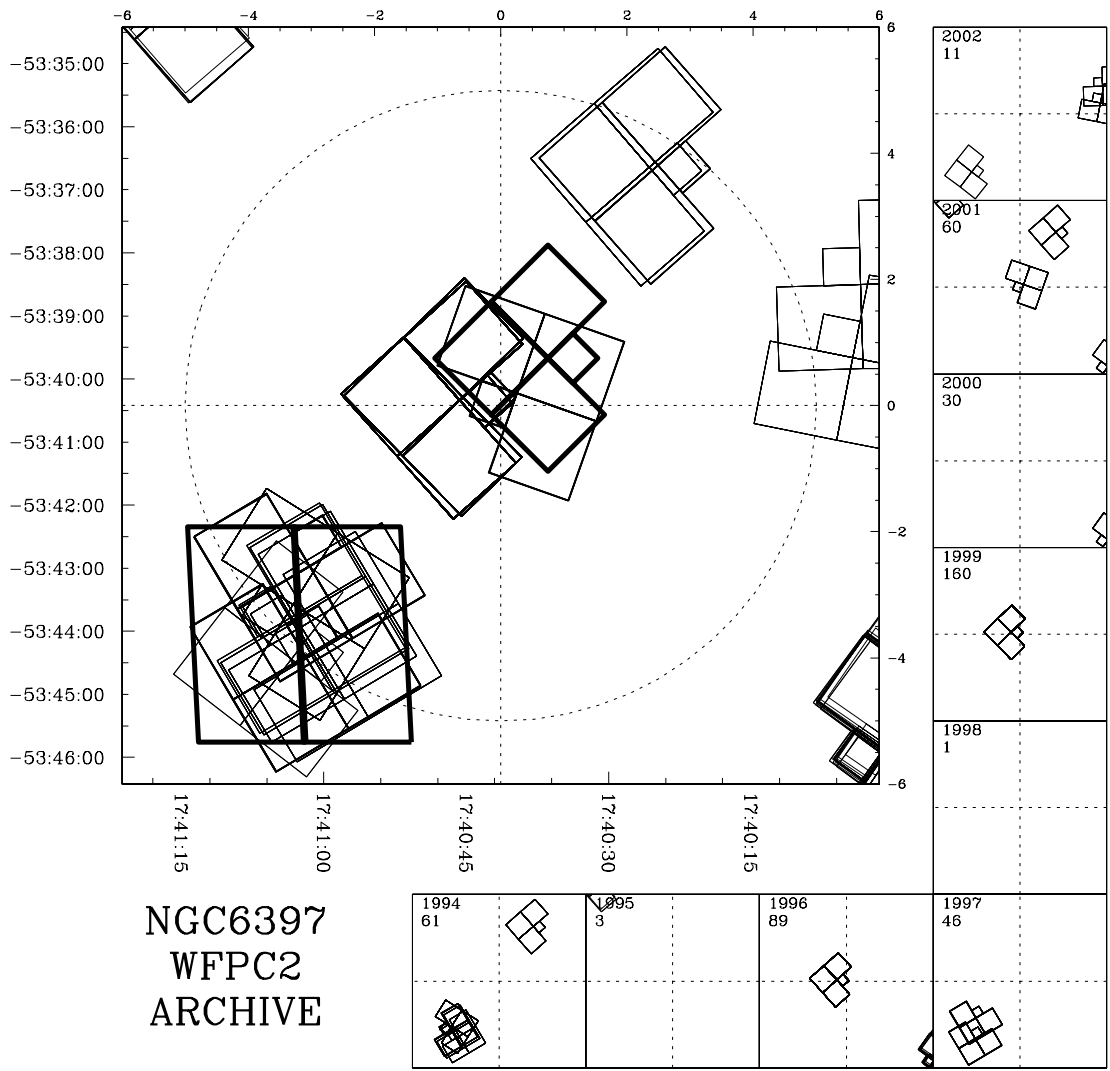}
\caption{Location of the field observed in NGC 6397. The cluster center is at (0,0) and RA and DEC are indicated (arc minutes from the cluster center shown along the top and right). The ACS field in which we observed is outlined
with the heavy solid lines (lower left).  Pre-existing WFPC2 fields are also indicated and were used to
proper motion clean the data.  We also exposed simultaneously with the WFPC2. The roll angle of the telescope was controlled so that the WFPC2
fell on the cluster core. }
\label{fig1}
\end{figure}

\begin{figure}
\plotone{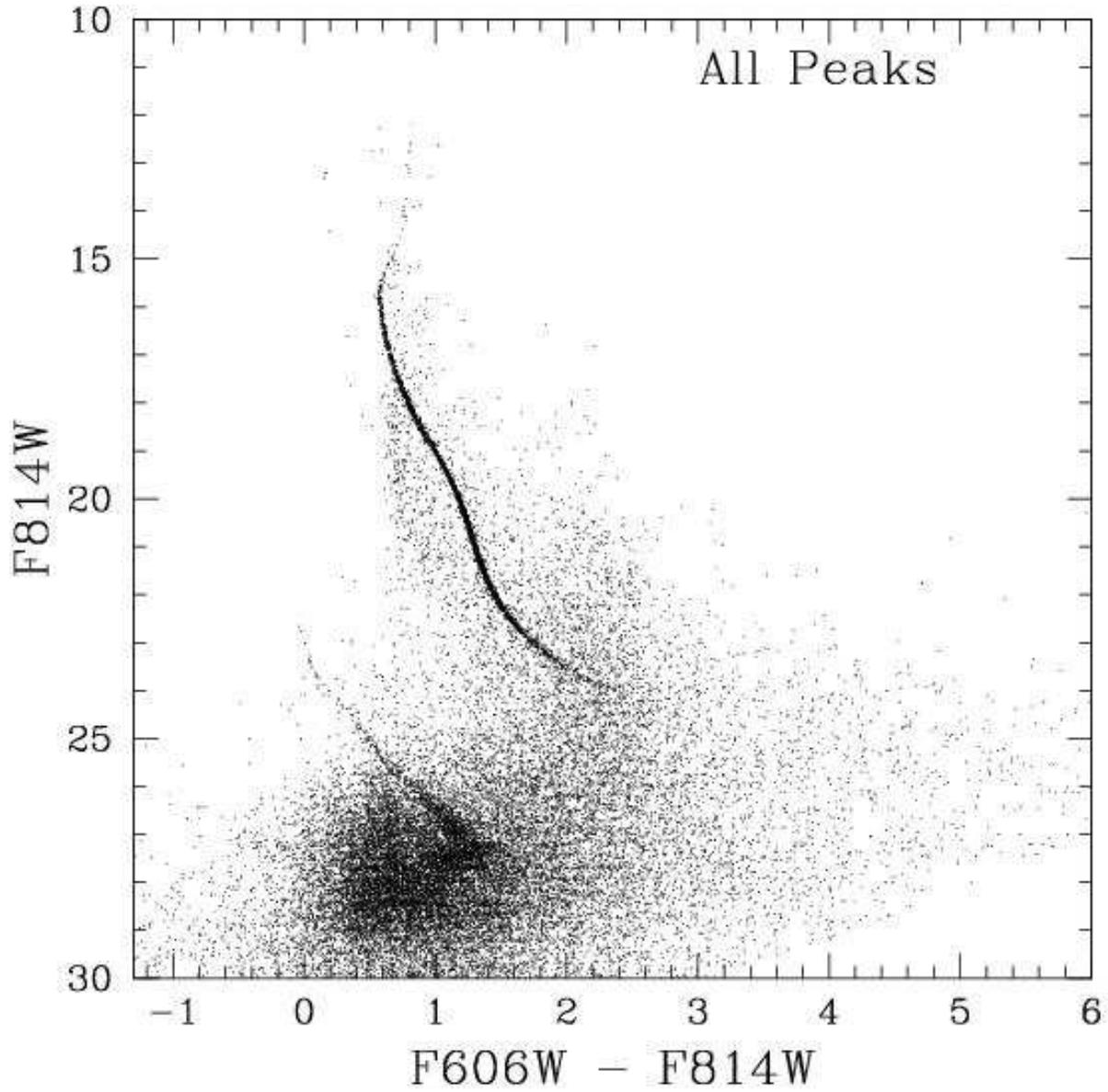}
\caption{ACS CMD in our NGC 6397 field located $5\arcmin$ SE (just under 2 half mass radii (Harris 1996)) from the cluster center.  This CMD contains all objects that contain a peak at the object's position in a minimum of 90/252 of the F814W images and are also the brightest peaks within 7.5 ACS pixels. There are a total of 48,785 objects in this CMD many of which are not stars.}
\label{fig2}
\end{figure}

\begin{figure}
\plotone{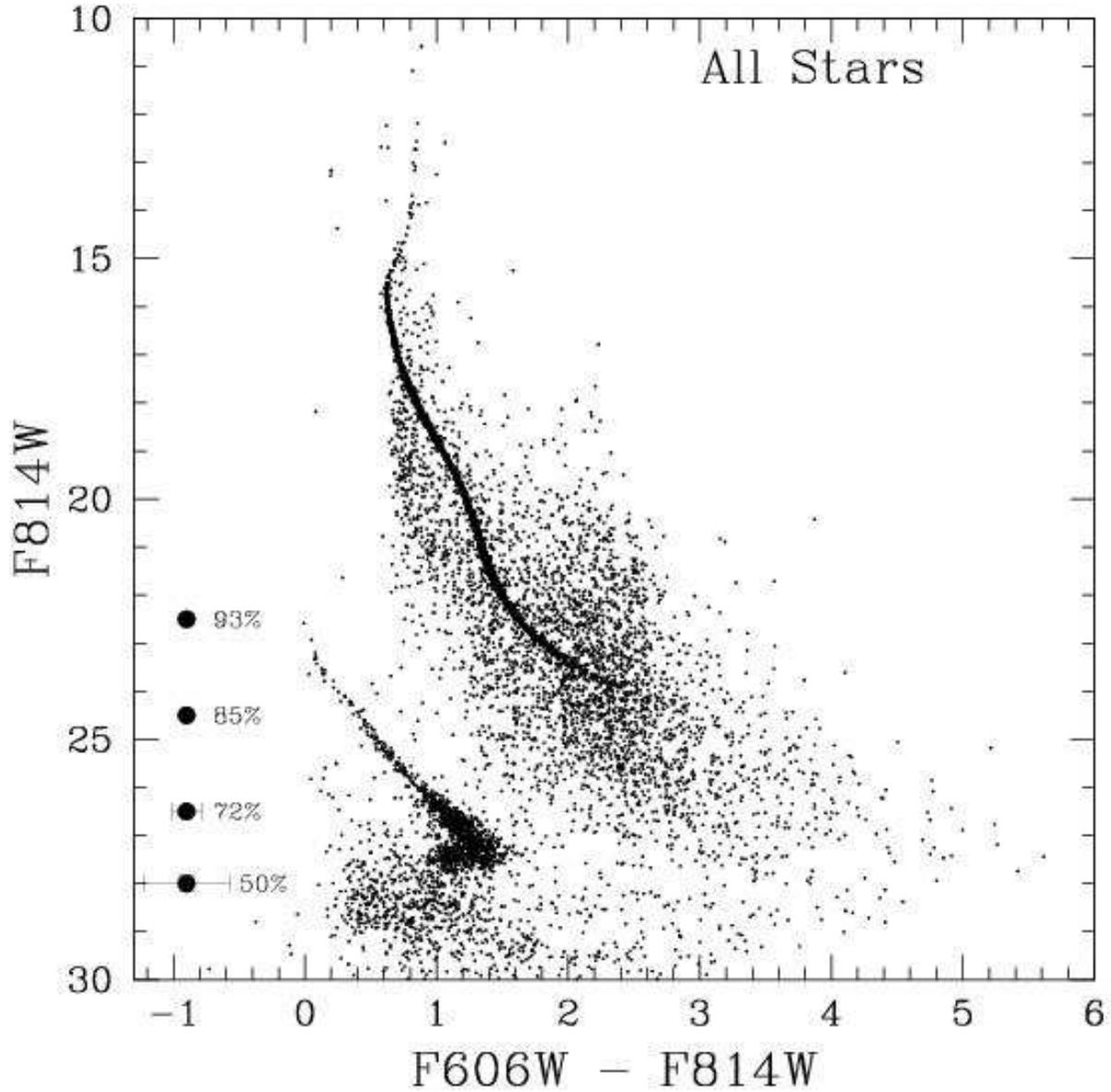}
\caption{This CMD contains all objects that pass our tests for separating stars from non-stellar objects. There are 8,537 stars in this diagram. The dots indicate the magnitude of the associated completeness for finding stars from our incompleteness tests, while the error bar is the $1\sigma$ photometric error in the F814W photometry at that magnitude.}
\label{fig3}
\end{figure}

\begin{figure}
\plotone{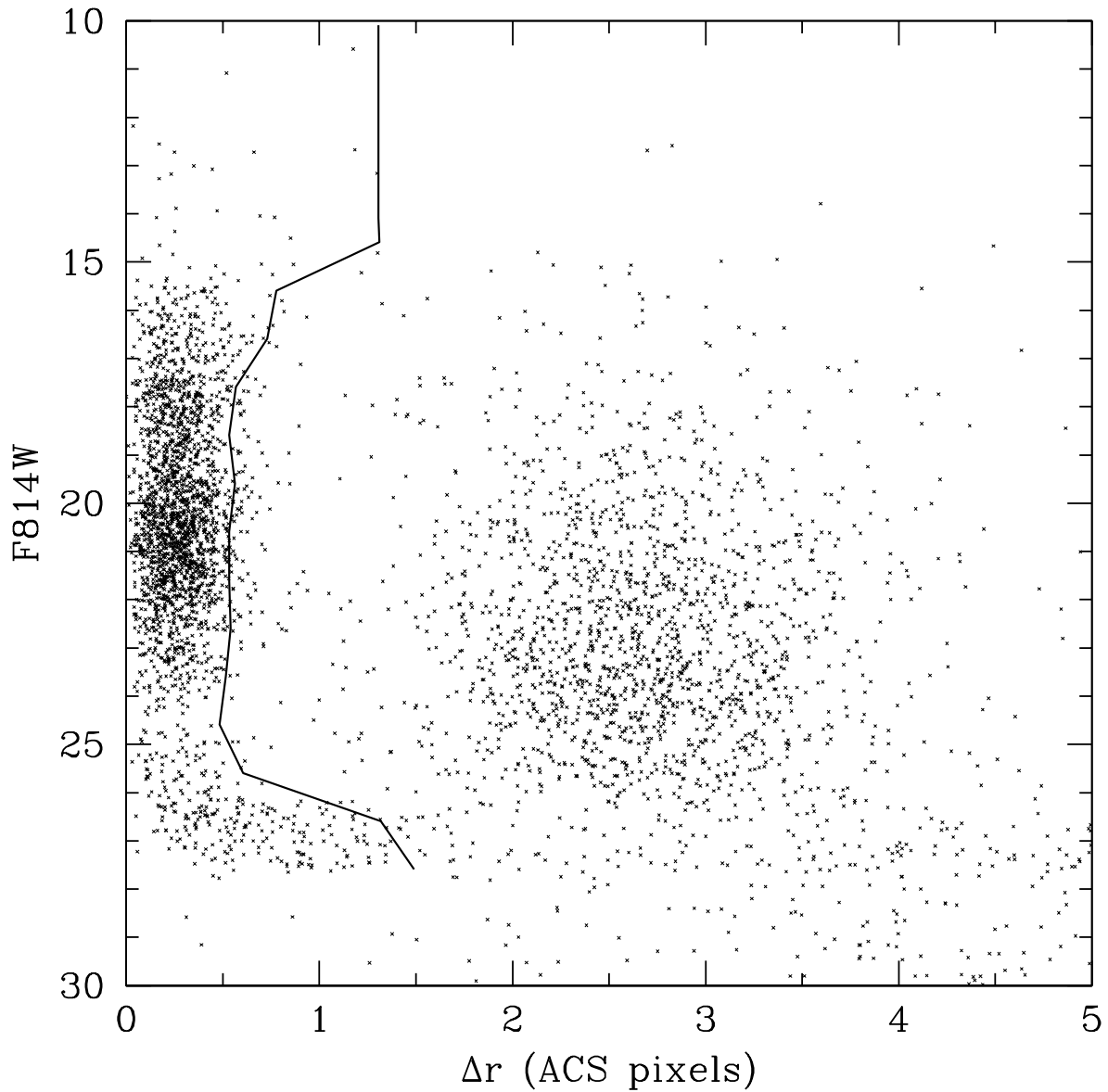}
\caption{ Total proper motion displacement of stars in our field
in ACS pixels over 10 years plotted against F814W magnitude. The proper motions
are zero-pointed on the cluster. The solid line is the $2\sigma$
error in the proper motion distribution at each magnitude. All stars to the left of this limit are taken to be cluster stars. The large errors at bright magnitudes are caused by stars that are saturated or nearly so.
The width of the distribution is a combination of measurement error and the internal proper motion
dispersion of the cluster. Note that proper motion cleaning for stars fainter than about $F814W = 28$ was not possible with these data. The limiting factor is the depth of the first epoch images.}
\label{fig4}
\end{figure}

\begin{figure}
\plotone{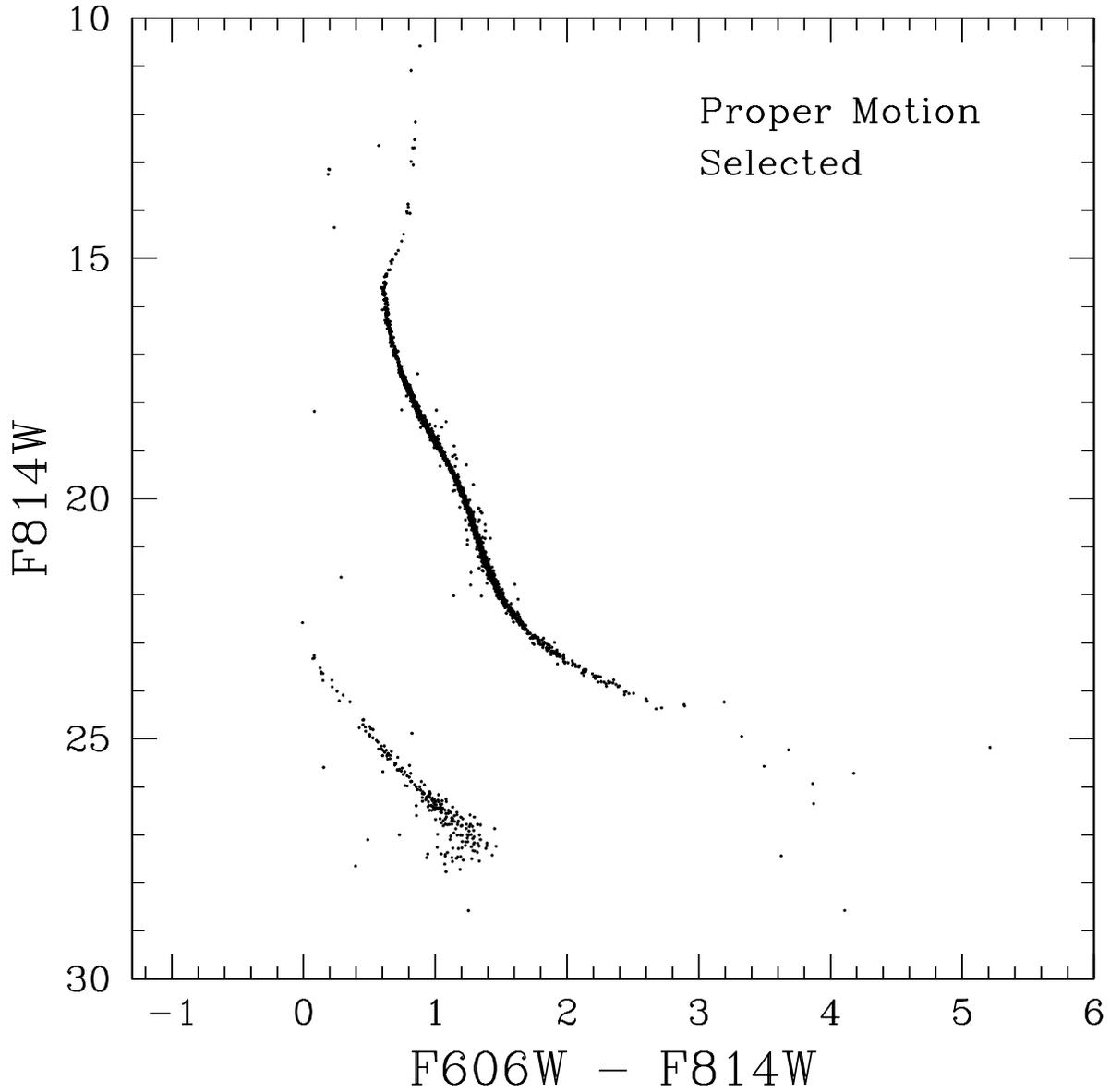}
\caption{ Proper motion cleaned CMD. All objects to the left of the solid line in Figure 4 are included in this diagram as well as a small number of proper motion selected objects from the short exposure frames so that
the bright MS and giant stars could be included.  Note that the ``blue hook'' in the white dwarf
region is preserved after proper motion cleaning and that the cluster MS extends to at least 
$F814W = 26$, $(F606W - F814W) = 4$.} 
\label{fig5}
\end{figure}

\begin{figure}
\plotone{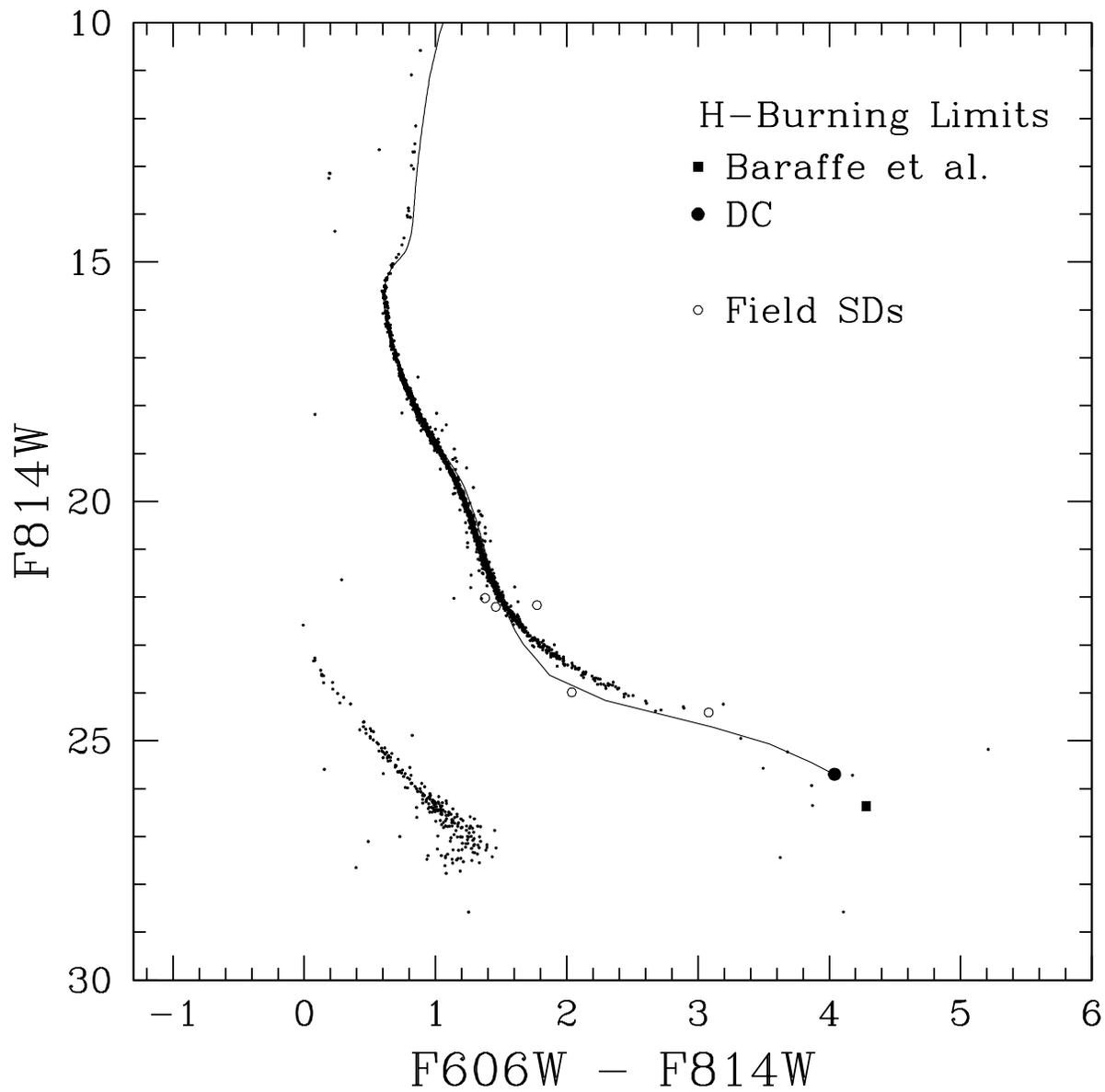}
\caption{ Same as Figure 5 except that we have added two theoretical points for the termination of the hydrogen burning sequence in this metal-poor cluster (Baraffe et al. 1997 and Dotter et al. 2007). Note that the data appear to terminate at about the magnitude predicted by these models. In addition we have included the 5 low metallicity field subdwarfs from Table 2 and an isochrone from Dotter which extends   to the hydrogen-burning limit.}
\label{fig6}
\end{figure}

\begin{figure}
\plotone{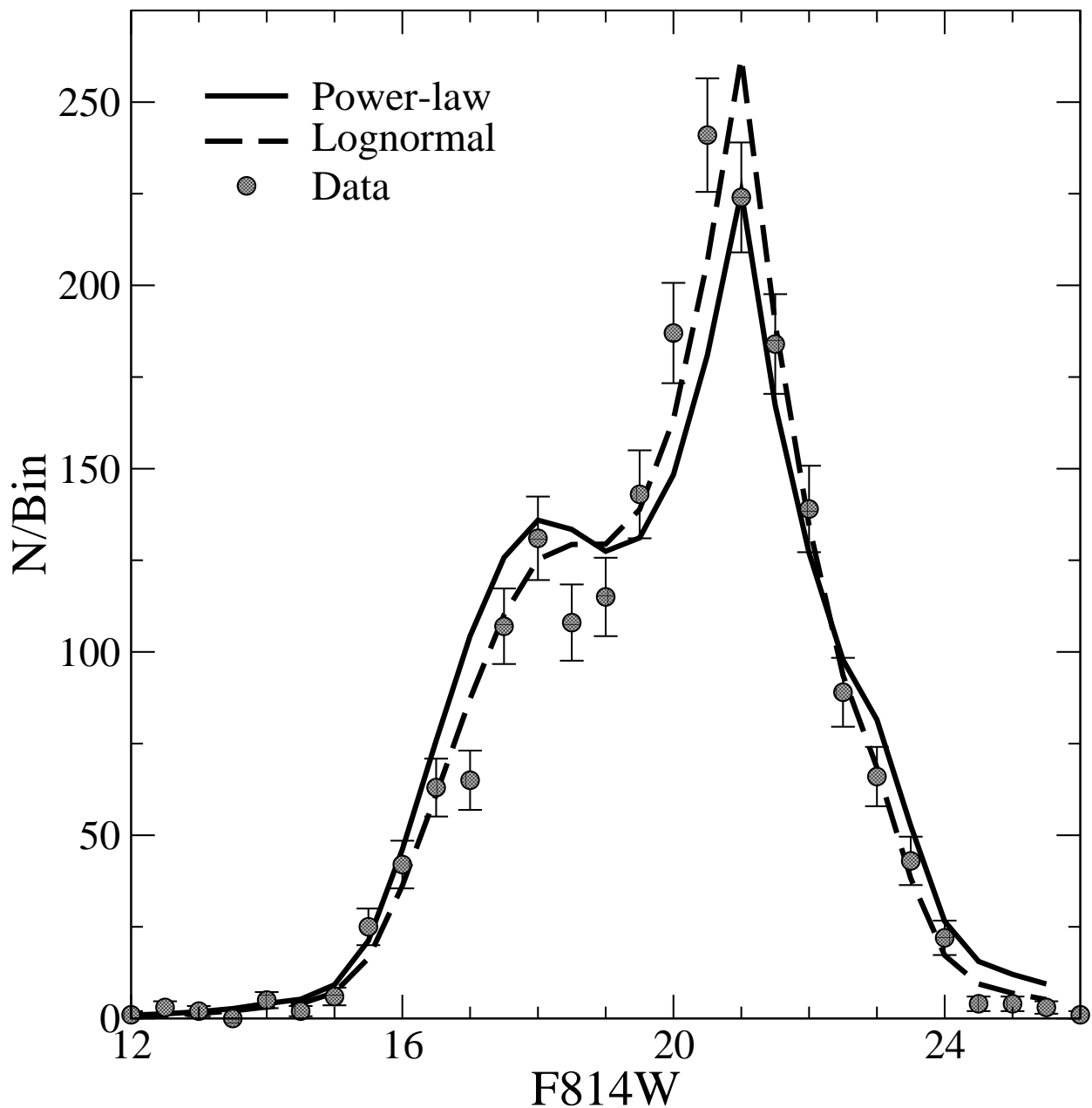}
\caption{NGC 6397  MS LF (circles with error bars) compared with the best fitting power-law and lognormal mass functions. The MF slope used to derive the power-law LF is $\alpha = 0.13$, while the best fitting lognormal function has $M_c = 0.27 M_{\odot}$ and $\sigma = 1.05$. Incompleteness corrections were applied to the theoretical functions.}
\label{fig7}
\end{figure}

\begin{figure}
\plotone{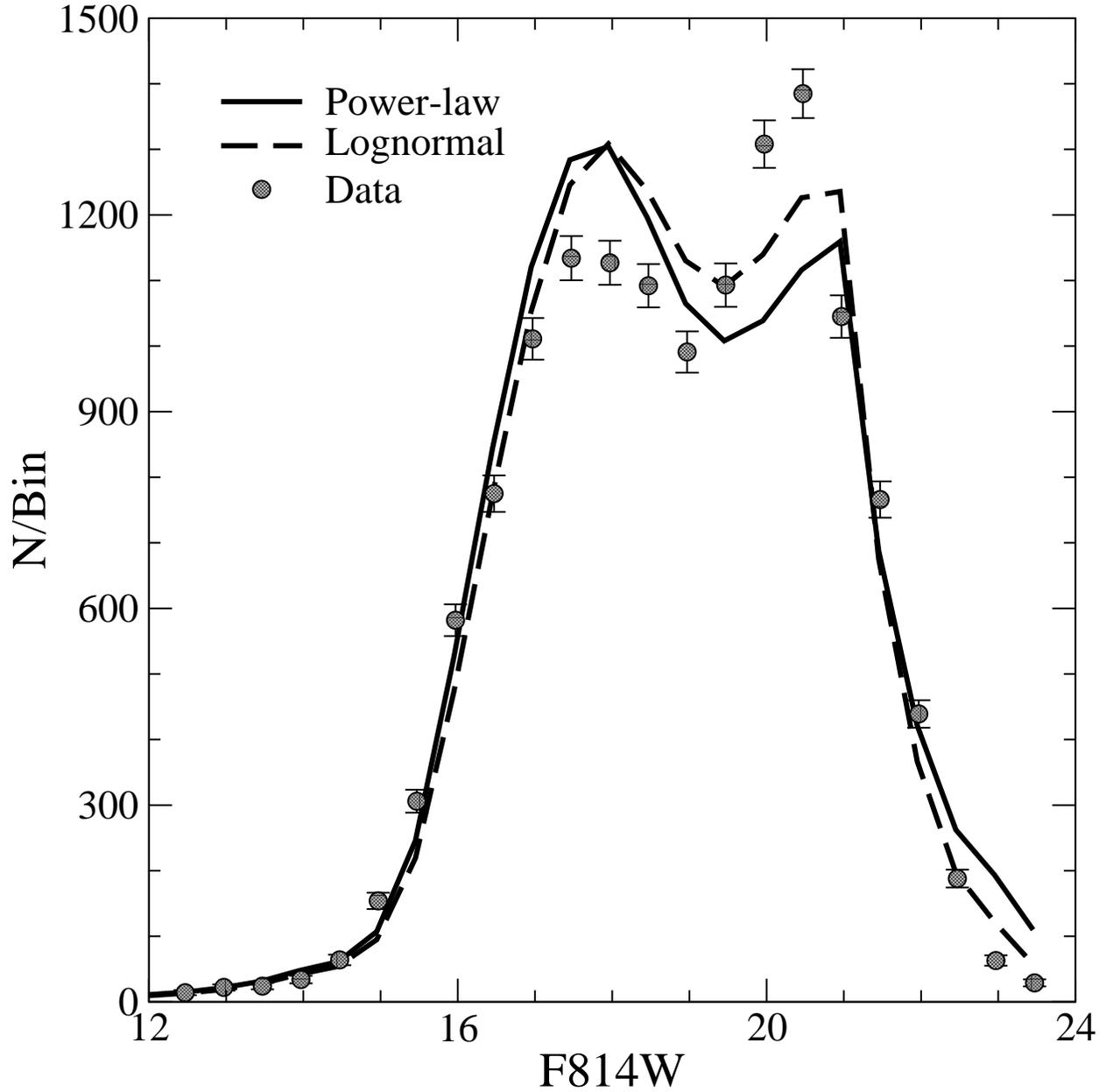}
\caption{ The NGC 6397 MS LF in the cluster core (circles with errors bars) compared
with the best fitting power-law and lognormal mass functions. The MF slope is $\alpha = -0.68$,
while the best fitting lognormal function has $M_c = 0.86 M_{\odot}$ and $\sigma = 1.05$. Incompleteness corrections were applied to the theoretical functions.}
\label{fig8}
\end{figure}

\begin{figure}
\plotone{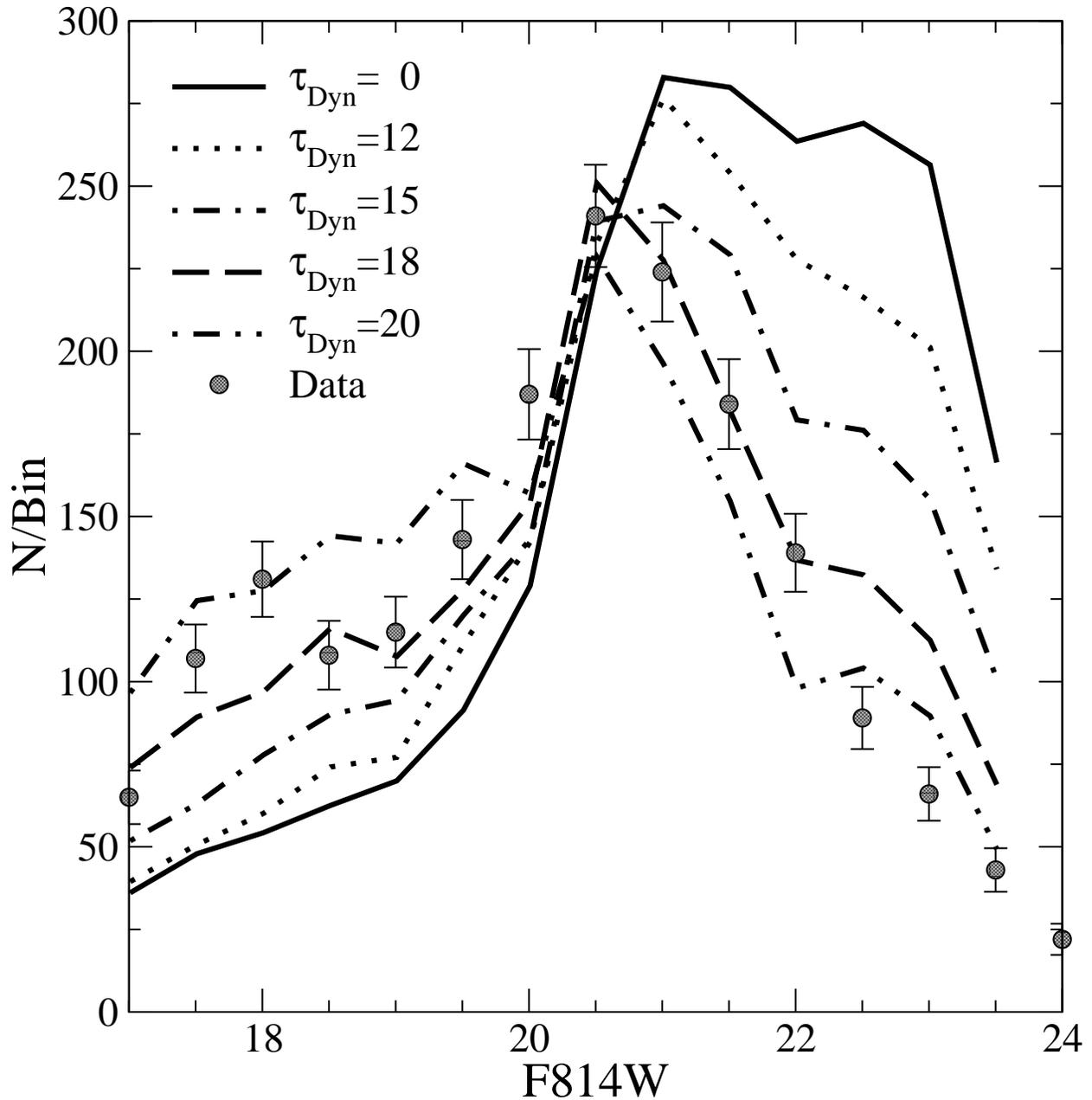}
\caption{Theoretical LFs for the ``simulated'' NGC 6397 cluster at various  dynamical ages for the outer field. The LFs are normalized so that they all have the same number of stars as the cluster ($1\,864$). The plot for $t_{Dyn} = 0$ is derived from the cluster initial mass function (IMF).}
\label{fig9}
\end{figure}

\begin{figure}
\plotone{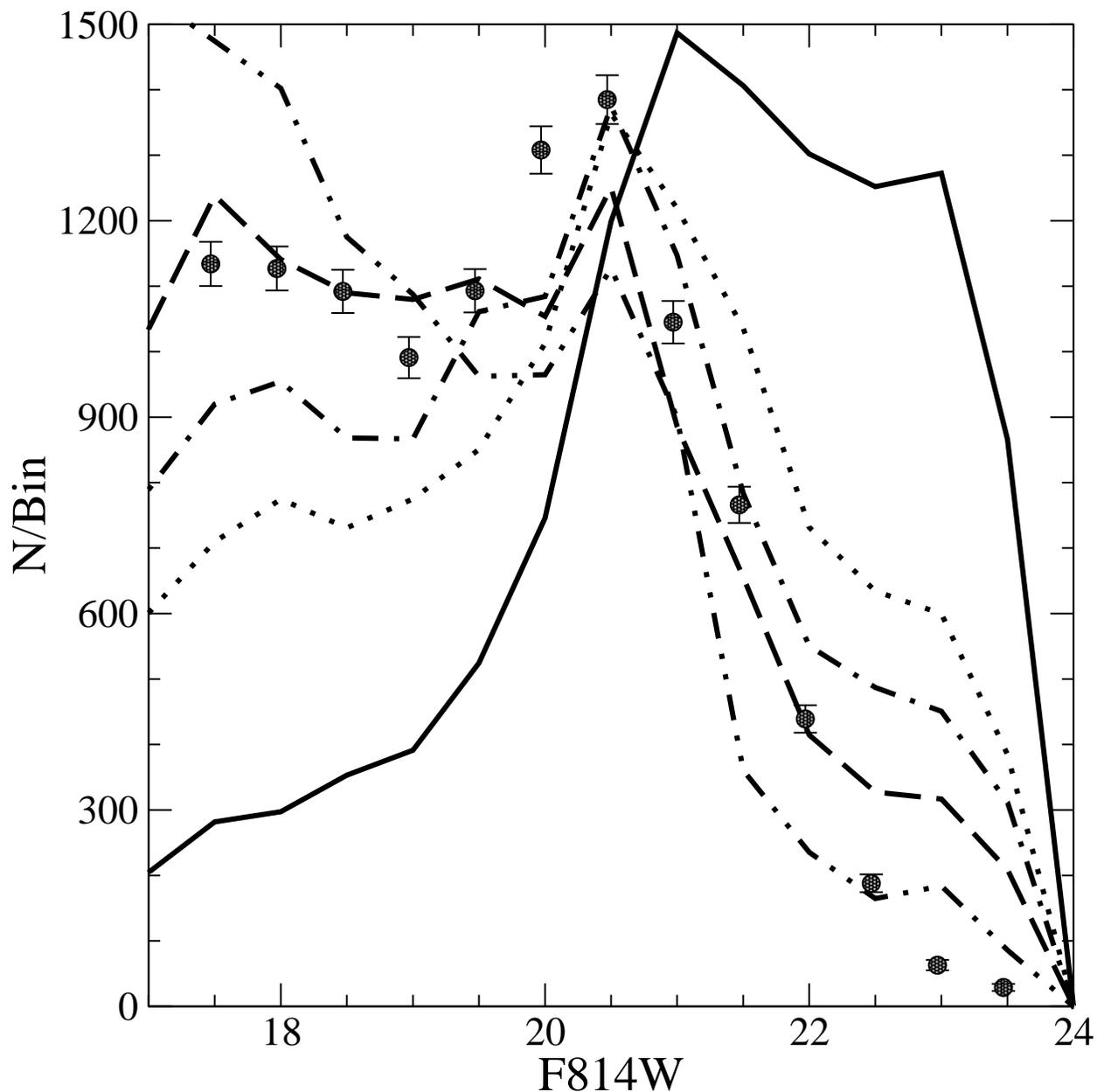}
\caption{Theoretical LFs for the ``simulated'' NGC 6397 cluster at various dynamical ages for the core field. The LFs are normalized so that they all have the same number of stars as the cluster ($13\,761$).  Because
of extreme crowding in the cluster core, stars within 0.1 half light radii ($\rm 15\arcsec$) were excluded from both the data and the simulation. The line types are the same as for Figure 9.}
\label{fig10}
\end{figure}

\begin{figure}
\plotone{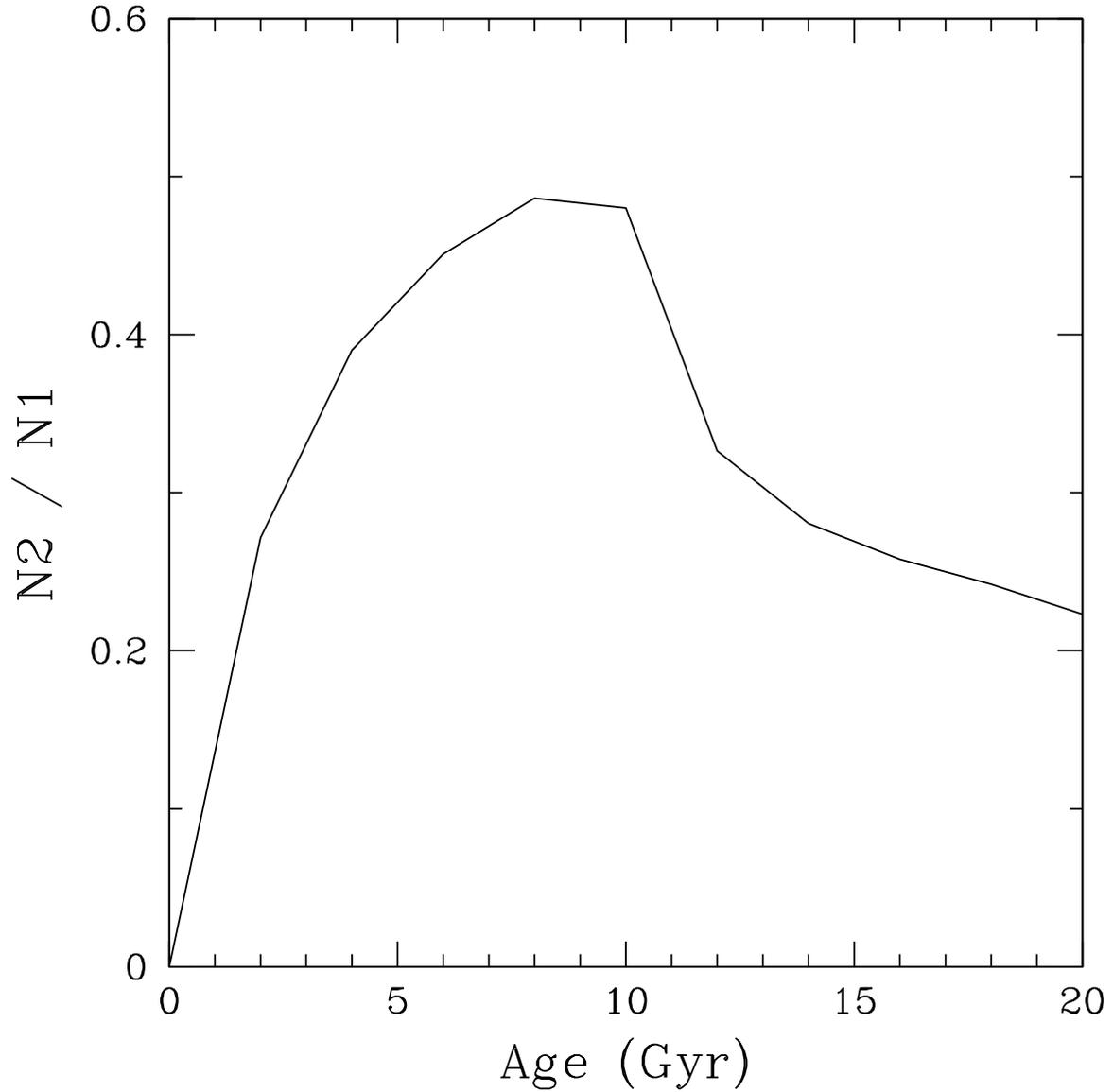}
\caption{ N-body model ratio in our field of the number of WDs (multiplied by the completeness correction so it matches the data): $N_2$ to the
number of MS stars in the mass range $ 0.5 - 0.7 M_{\odot}: N_1$. The plotted ratio includes the effects of both stellar
and dynamical evolution.}
\label{fig11}
\end{figure}

\begin{figure}
\plotone{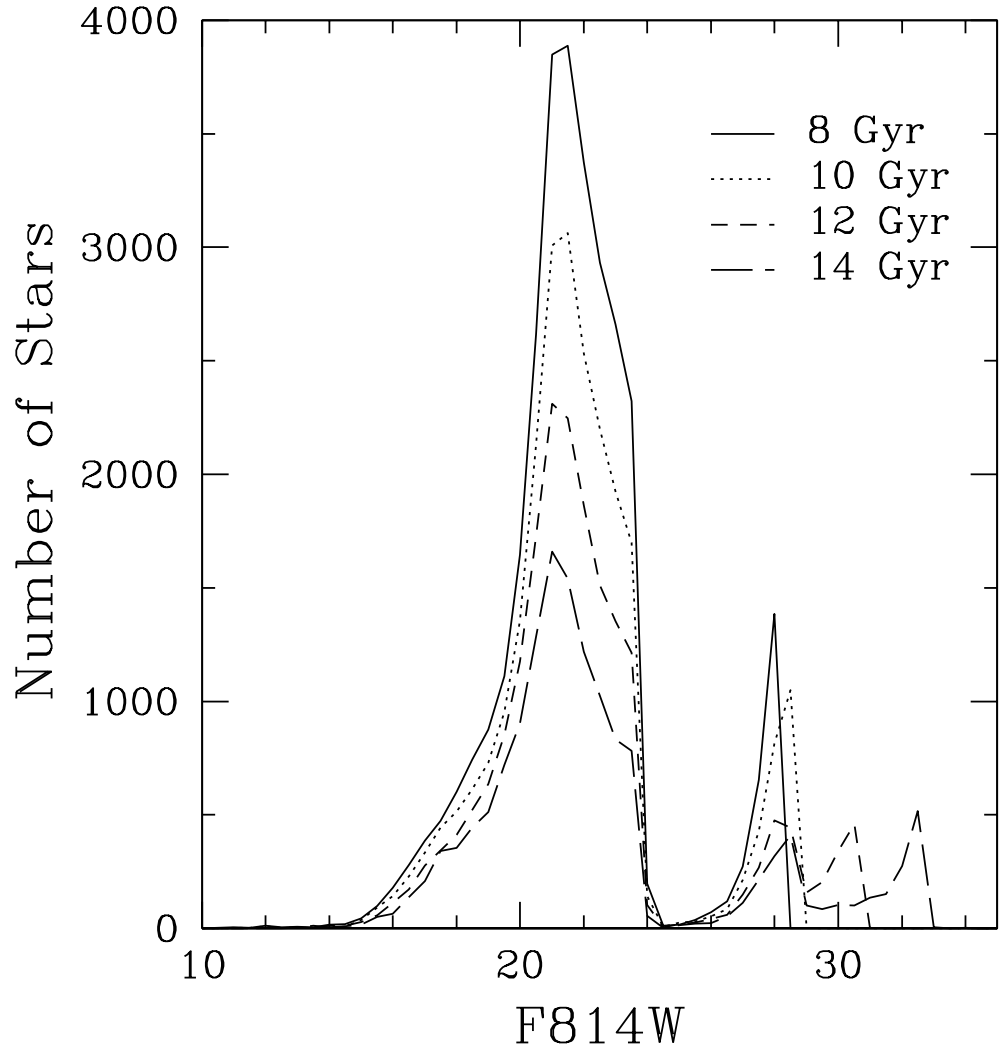}
\caption{ Simulated LFs at various ages in our field of NGC 6397. This is an example of a LF that JWST could potentially produce with a very deep exposure in this cluster and with a second epoch for proper motion cleaning. Images in only a single color would be required thus substantially reducing observing time.}
\label{fig12}
\end{figure}

\clearpage

 \begin{deluxetable}{lllcrr}
\tablewidth{0pc}
\tablecaption{Photometry of Proper Motion Selected NGC 6397 Cluster Stars \label{table1}}
\tablehead{
\colhead{$x$}  & \colhead{$y$}  & \colhead{$F814W$}  & \colhead{$(F606W - F814W)$}  & \colhead{$dx$}  & \colhead{$dy$}
 }
 \startdata

       82.2    &   787.2  &     17.12 &     0.7258  &     0.02  &     -0.25  \\
       85.62    &   1003  &     17.58  &    0.7834 &    0.38  &     -0.39   \\
       98.81    &    1154 &      21.11 &      1.359  &    0.19  &      -0.2  \\
       104.7    &   956.7  &      21.7  &     1.456  &     0.31   &     0.28  \\
       106.4   &    972.1  &     20.88   &     1.34   &    -0.38   &    -0.06  \\

 \enddata
\tablerefs{$x$ and $y$ are the pixel values for the stars  as they appear on the reference frame
 j97101bbq.flt. $F814W$ and ($ F606W - F814W$)
are the VEGAMAGS of the stars calibrated according to Sirianni et al. (2005), and $dx$ and $dy$ are the $x$ and $y$ pixel shifts in the positions of the stars over a 10 year period with respect to the cluster.  The ACS pixels are $0.05\arcsec$ on a side. The few entries listed 
in this Table are to illustrate the format, the complete Table can be found via the link to the machine-readable version above.}
 \end{deluxetable}
\clearpage


\begin{deluxetable}{cccccc}
\tablecolumns{6}
\tablewidth{0pc}
\tablecaption{Low Luminosity Metal-Poor Subdwarfs in the Halo \label{table2}}
\tablehead{
\colhead{Star}  & \colhead{$V$} & \colhead{$I_C$} & \colhead{$d$ (pc)} & \colhead{$F814W$}
 & \colhead {$(F606W - F814W)$}
 }
 \startdata
 
LHS 169 & 14.13 & 12.41  & 32.4  & 22.20   & 1.46  \\
LHS 377  & 18.39 & 14.91 &  35.2  & 24.41  & 3.08 \\
LHS 407  & 16.57 & 14.18 &  31.7  &  23.99 &  2.04\\
LHS 522  & 14.15 & 12.53 &  37.3  &   22.02 &  1.38 \\
LHS 541 & 16.46 & 14.37 & 80.6   &  22.17 &1.77 \\
 \enddata
\tablerefs{Photometry and distances taken from Gizis \& Reid (2000); metallicities of all these objects are $ < -1.5$
and are taken from Gizis (1997), Phan-Boa \& Bessell (2006) and Woolf \& Wallerstein (2006), transformation equations from Sirianni et al. (2005); ACS magnitudes are those the star would have if it were in NGC 6397.}
\end{deluxetable}

\begin{deluxetable}{lllr}
\tablewidth{0pc}
\tablecaption{NGC 6397 Physical Parameters \label{table3}}
 \tablehead{
\colhead{Parameter} & \colhead{Value} & \colhead{Comment} 
 & \colhead{Reference} 
 }
\startdata

distance (kpc) & 2.53 $\pm$ 0.03 & Subdwarf fit  in $B$ and $V$& 1\\
\phm{distance} & 2.58 $\pm$ 0.04 & Subdwarf fit in $b$ and $y$ & 1 \\
\phm{distance} & 2.67 $\pm$ 0.15 & Subdwarf fit & 2 \\
 true distance modulus & 12.07 $\pm$ 0.06 & as above & 1,2 \\
true distance modulus & 12.03 $\pm$ 0.06 & WDs & 8\\
\\
$E(B-V)$  & 0.18 $\pm$ 0.01       & Subdwarf fit & 1,2 \\
 $E(F606W - F814W)$ & 0.20 $\pm$ 0.03& WD fitting & 8\\
 $A_{ V}$     & 0.56  & Model extinction curves & 3 \\
$A_{  F606W}$     & 0.51  & HST Transformations& 4 \\
A$_{ F814W}$     & 0.33  & HST Transformations & 4 \\
$  [Fe/H]$;$  [\alpha/Fe]$   & $-1.82$ $\pm$ 0.04; +0.3 & Stromgren Photometry & 5 \\
$  [Fe/H]$;$  [\alpha/Fe]$   & $-2.03$ $\pm$ 0.05; +0.34 & Spectroscopy & 1 \\
$  [Fe/H]$   & $-2.02$ $\pm$ 0.07& Spectroscopy & 6 \\

Age (Gyr) & 11.43 $\pm0.46 (2\sigma)$ & WD Cooling & 8\\
Age (Gyr) & 12 $\pm 1(1\sigma)$ & Main Sequence Fitting& 1\\

\\
core radius ($r_c$)     & 3$\arcsec$   & King model  & 7 \\
half light radius ($r_h$)     & 174$\arcsec$   & King model  & 7\\
concentration ($c$) & 2.5 (collapsed core)      & King model  & 7\\
half mass relax. time ($t_{rh}$) & 2.9$\times 10^8$ yr. 
    & Spitzer formula &  9\\
central relax. time ($t_{rc}$)
    & 7.9$\times 10^4$ yr.  & Spitzer formula & 9\\
\enddata
\tablerefs{(1) Gratton et al. 2003; (2) Reid \& Gizis 1998; 
(3) Cardelli et al. 1989; (4) Sirianni et al. 2005; (5) Twarog \& Twarog 2000; 
(6) Kraft \& Ivans 2003; (7) Trager et al. 1995; (8) Hansen et al. 2007;  (9) Harris 1996.}
\end{deluxetable}

\clearpage

\begin{deluxetable}{cccc}
\tablecolumns{4}
\tablewidth{0pc}
\tablecaption{Completeness Fractions \label{table4}}
\tablehead{
\colhead{$F814W$}  & \colhead{Completeness} & \colhead{$F814W$} & \colhead{Completeness}  
 }
 \startdata
 
22.214 & 92.18 & 26.164  & 75.62 \\
22.664  & 92.27 & 26.414 &  72.55 \\
22.914  & 91.96 & 26.664 &   71.84 \\
23.164  & 91.93 & 26.914 &  70.31   \\
23.414 & 90.24 & 27.164  & 66.17  \\
23.664  & 89.22 & 27.414 &   58.97  \\
23.914  & 89.71 & 27.664 &  55.53  \\
24.164  & 88.01& 27.914 &   48.93\\
24.414 & 85.87 & 28.164  &   36.86\\
24.664  & 84.95 & 28.414 &   21.22 \\
24.914  & 84.72 & 28.664 &   8.01 \\
25.164  & 81.97 & 28.914 &  2.20\\ 
25.414 & 79.40 & 29.164  &   0.33\\
25.664  & 80.19 & 29.414 &   0.07 \\
25.914  & 77.96 & 29.664 &   0.00\\
 \enddata
 \end{deluxetable}

\clearpage

   \end{document}